\RequirePackage{color}

\documentclass[11pt,a4paper]{article}
\pdfoutput=1
\usepackage{jheppub}

\usepackage{latexsym,amsmath,amsfonts,amssymb}
\usepackage{graphicx}
\usepackage{multirow}
\usepackage{wrapfig}

\newcommand{\ket}[1]{\mathchoice{
    {\left|{#1}\right\rangle}}{|{#1}\rangle}{|{#1}\rangle}{|{#1}\rangle}}

\newcommand{\be}{\begin{equation}}
\newcommand{\ee}{\end{equation}}

\newcommand{\barray}{\begin{eqnarray}}
\newcommand{\earray}{\end{eqnarray}}

\newcommand{\cC}{\mathcal{C}}

\newcommand{\cI}{\mathcal{I}}

\newcommand{\cM}{\mathcal{M}}
\newcommand{\cN}{\mathcal{N}}
\newcommand{\cO}{\mathcal{O}}

\newcommand{\cR}{\mathcal{R}}

\newcommand{\bC}{\mathbb{C}}

\newcommand{\bZ}{\mathbb{Z}}

\newcommand{\lp}{\left(}
\newcommand{\rp}{\right)}
\newcommand{\hA}{\hat{A}}
\newcommand{\pe}{\text{P.E.}}

\newcommand{\beq}{\begin{equation}}
\newcommand{\eeq}{\end{equation}}

\DeclareMathOperator{\Tr}{Tr}

\def\drawbox#1#2{\hrule height#2pt 
        \hbox{\vrule width#2pt height#1pt \kern#1pt \vrule width#2pt}
              \hrule height#2pt}
\def\Asym#1#2{\vcenter{\vbox{\drawbox{#1}{#2}
              \kern-#2pt       % line up boxes
              \drawbox{#1}{#2}}}}

\title{Chiral Ring Generating Functions \& Branches of Moduli Space}

\author{James McGrane, Sanjaye Ramgoolam, Brian Wecht}
  
\affiliation{Centre for Research in String Theory\\
Queen Mary University of London\\
London E1 4NS\\
United Kingdom}

\abstract{We consider the worldvolume theory of $N$ D3-branes transverse to  various non-compact Calabi-Yau spaces, and describe subtleties in the counting of chiral primary operators in such theories due to the presence of multiple branches of moduli space. Extra branches, beyond those directly related to the transverse geometry, result in additional terms in the generating functions for single- and multi-trace operators. Ideals in the $N=1$ chiral ring correspond to various branches 
and, in the large $N$ limit, the operator counting reveals a product of Fock spaces, including the Fock space of bosons on the space transverse to the branes.  
}

\preprint{QMUL-PH-15-10}

\begin{document}

\maketitle

%%%%%%%%%%%%%%%%%%%%%%%%%%%%%%%%%%%%%%%%%%%%%%%%%%%%%%%%%%%%%%%%%%
\section{Introduction}
\label{sec:intro}
%%%%%%%%%%%%%%%%%%%%%%%%%%%%%%%%%%%%%%%%%%%%%%%%%%%%%%%%%%%%%%%%%%
%%%%%%%%%%%%%%%%%%%%%%%%%%%%%%%%%%%%%%%%%%%%%%%%%%%%%%%%%%%%%%%%%%

The study of D3-branes transverse to conical non-compact  Calabi-Yau spaces has a long and storied past. As the transverse geometry has increased in complexity from $\bC^3$ \cite{Witten:1995im} to the conifold \cite{Klebanov:1998hh} to orbifold singularities \cite{Douglas:1996sw,Morrison:1998cs} and beyond \cite{Beasley:1999uz, Benvenuti:2004dy, Franco:2005sm}, our understanding of the related worldvolume theories, as well as the techniques used to study them, has increased dramatically. The motivations for these studies have ranged across a variety of themes: 
 the fundamentals of D-brane physics, matrix models \cite{Banks:1996vh}, brane-engineering of gauge theory dynamics, geometric engineering and reverse geometric engineering \cite{Berenstein:2002ge}, and AdS/CFT \cite{Maldacena:1997re, Witten:1998qj, Gubser:1998bc}. Nevertheless, despite so many years of study, many interesting and important questions about these theories remain.

A particularly interesting part of any $ \cN =1$ supersymmetric theory is the set of chiral gauge-invariant operators. These operators are annihilated by the supersymmetry generators of one chirality,  $ \bar Q $, and are usefully considered modulo an equivalence relation where commutators with $ \bar Q $, i.e. $\bar Q $-exact operators,  are set to zero.  With this equivalence, derivatives can be set to zero. The ring formed by these operators, called the chiral ring, will play a pivotal role in the current work. In SCFTs, chiral primary operators (the lowest weight states in their representation of the conformal group)  
can be chosen as representatives of the  chiral ring equivalence classes. We will focus on the chiral ring operators which are 
constructed from matter multiplets. 

One particularly interesting question is how to derive the spectrum of chiral primary operators. For D3-branes at the tip of a Calabi-Yau cone, the dimensions of such operators can be computed with $a$-maximisation \cite{Intriligator:2003jj} (or, on the geometry side, $Z$-minimisation \cite{Martelli:2006yb}). Additionally, the dimension $\Delta$ of two gauge-invariant chiral primary operators $\mathcal{O}_1$ and $\mathcal{O}_2$ are additive in the sense that $\Delta(\mathcal{O}_1 \mathcal{O}_2) = \Delta (\mathcal{O}_1) + \Delta( \mathcal{O}_2)$. A central question about such operators is then how many there are with a given dimension. This counting has been achieved in many theories thanks to the ``plethystic program" of \cite{Benvenuti:2006qr,Feng:2007ur}. In many cases of interest, 
the theories have a number of $U(1)$ global symmetries. A basis of the chiral ring can be formed from operators with definite 
charges, and generating functions can be defined for this refined counting.

A closely related object of interest in a supersymmetric gauge theory is the moduli space of constant (space-time independent) 
zero energy configurations of  the scalar  matter fields. 
Since the energy is a sum of squares of F- and D-terms, these configurations 
solve D- and F-term equations. Vacuum expectation values of gauge-invariant chiral ring operators can be used to parameterize 
the  moduli space. As a result, the chiral ring is expected to be the ring of holomorphic polynomial functions on the moduli space  (see e.g. \cite{Kutasov:1995ss,Luty:1995sd}).  This connection between the space and the ring is of the form one encounters in algebraic geometry, where the  study of ideals in 
the ring is an important part of the story.

It is natural to  interpret the moduli space of the worldvolume theory of a stack of D3-branes as the transverse geometry. For example, in the case of a single, flat D3-brane in $ {\mathbb{R}}^{ 9,1} $, the three (uncharged) chiral superfields in the worldvolume theory naturally correspond to the coordinates on the transverse space $\bC^3$. For multiple branes, the moduli space is expected to 
be a symmetric product $ Sym^N ( X )$, where $X$ is the transverse space and $N$ is the number of branes; similarly, the gauge group of the worldvolume gauge theory is a product of $U(N)$ factors. The ring of functions on the symmetric product 
corresponds to bosonic wavefunctions of an $N$-particle system on $X$. 
The explicit demonstration for $ \bC^3$ is in \cite{Witten:1995im} 
and for the conifold in \cite{Klebanov:1998hh}; the moduli spaces for orbifold theories are also considered in \cite{Douglas:1996sw,Douglas:1997de}.
The appearance of  symmetric products plays an important role in matrix theory \cite{Banks:1996vh}
as well as reverse geometric engineering \cite{Berenstein:2002ge}.   In the large $N$ limit, the ring of functions on $ Sym^N ( X ) $ 
can be mapped to a Fock space of states obtained by acting on a vacuum with oscillators, one for each holomorphic monomial 
function on the space $X$. The emergence of Fock spaces at large $N$  is central to the  AdS/CFT correspondence. 
 On the AdS  side, the Fock spaces arise from multi-particle states obtained from Kaluza-Klein  reduction on the base of the cone transverse to the 3-branes. The counting of Fock space states is related to the counting of single particle states by the plethystic exponential 
 and, as such,  this   has  played an important role in the plethystic program \cite{Feng:2007ur,Benvenuti:2006qr, Hanany:2006uc}. The problem of counting chiral operators in M2-brane world-volume gauge theories transverse to orbifold geometries was also considered in \cite{Hanany:2008qc}. A similar problem of calculating the superconformal index \cite{Kinney:2005ej,Romelsberger:2005eg} for D3-brane worldvolume gauge theories has been completed for the transverse geometries $\bC^3$ \cite{Kinney:2005ej}, the conifold \cite{Nakayama:2006ur}, and other orbifold theories \cite{Nakayama:2005mf}.

The goal of the present work is to explore these relationships between chiral rings, moduli spaces, large $N$ Fock spaces,
 and the transverse geometry in various  examples of D3-branes transverse to non-compact Calabi-Yau spaces. 
We will pay particular attention to the fact that,  even in very simple cases such as $ \bC^3/\bZ_2$, the 
existence of multiple branches of the moduli space brings additional subtleties to the web of inter-relations
linking gauge theory combinatorics to geometry.  Along the way, we will also examine  the close relation observed \cite{Benvenuti:2006qr} between 
the chiral ring of the $U(1)$ theory and the single trace operators in the large $N$ theory (throughout this work, we will refer 
to the large $N$ theory as the $ U(\infty)$ theory). In this paper we will only consider mesonic operators since the gauge group will always be a product of unitary gauge groups. The baryonic branch has been considered for similar theories with special unitary gauge groups in \cite{Forcella:2007wk,Butti:2007jv}.

The outline of the remainder of this paper is as follows. In Section 2, we review some basic technology for quiver theories, chiral rings, and the generating functions that count chiral primaries. The remaining sections then consider a variety of examples in increasing order of complexity: $\cN=4$ (Section 3), the conifold (Section 4), $\bC^3/\bZ_2$ (Section 5), $\bC^3/\bZ_n$ (Section 6), and $\bC^3/ \hA_n$ (Section 7), where $\hA_n$ is the order $n$ cyclic subgroup of $SU(2)$. Finally, in Section 8, we briefly conclude, and various details are relegated to appendices.

%%%%%%%%%%%%%%%%%%%%%%%%%%%%%%%%%%%%%%%%%%%%%%%%%%%%%%%%%%%%%%%%%%%
\section{Review}
\label{sec:review}
%%%%%%%%%%%%%%%%%%%%%%%%%%%%%%%%%%%%%%%%%%%%%%%%%%%%%%%%%%%%%%%%%%%
%%%%%%%%%%%%%%%%%%%%%%%%%%%%%%%%%%%%%%%%%%%%%%%%%%%%%%%%%%%%%%%%%%%
Throughout this paper we will be looking at gauge theories that live on the worldvolume of a flat stack of D3-branes with a 3-complex dimensional transverse space. Our main goal in this work is to describe the relationships between the following objects:
\begin{itemize}
\item The space transverse to the D3-branes
\item The moduli space of the world-volume theory
\item The $U(1)$ ({\it i.e.}, single-brane) chiral ring of the world-volume theory
\item The set of single-trace operators of the world-volume theory
\item The ring of multi-trace operators of the world-volume theory
\item The generating functions for counting operators in these sets/rings
\end{itemize}
In this section, we review these concepts.

%%%%%%%%%%%%%%%%%%%%%%%%%%%%%%%%%%%%%%%%%%%%%%%%%%%%%%%%%%%%%%%%%%%
\subsection{Moduli Space}
\label{subsec:revmodsp}
%%%%%%%%%%%%%%%%%%%%%%%%%%%%%%%%%%%%%%%%%%%%%%%%%%%%%%%%%%%%%%%%%%%
%%%%%%%%%%%%%%%%%%%%%%%%%%%%%%%%%%%%%%%%%%%%%%%%%%%%%%%%%%%%%%%%%%%
It is common for supersymmetric theories to have a moduli space of supersymmetric vacua. For a SUSY gauge theory with gauge group $G$, three different but equivalent ways of finding the classical moduli space are:
\begin{enumerate}
\item Solve the D- and F-term relations modulo $G$ transformations.
\item Solve the F-term relations modulo $G_{\bC}$ transformations, where $G_{\bC}$ is the complexified gauge group.
\item Find the holomorphic gauge-invariant monomials modulo algebraic relations.
\end{enumerate}
For a review, see {\it e.g.~}\cite{Argyresnotes, Luty:1995sd}. Throughout the present work, we will use methods 1 and 3. In the context of the moduli space, we will be talking about the ring of holomorphic polynomials on the space, so we now introduce some basic ideas in ring theory.
%%%%%%%%%%%%%%%%%%%%%%%%%%%%%%%%%%%%%%%%%%%%%%%%%%%%%%%%%%%%%%%%%%%
\subsection{Rings, Ideals, and Quotient Rings}
\label{subsec:revriqr}
%%%%%%%%%%%%%%%%%%%%%%%%%%%%%%%%%%%%%%%%%%%%%%%%%%%%%%%%%%%%%%%%%%%
%%%%%%%%%%%%%%%%%%%%%%%%%%%%%%%%%%%%%%%%%%%%%%%%%%%%%%%%%%%%%%%%%%%
A ring $\cR$ is a set of elements with two binary operations: addition and multiplication. The ring is an abelian group under addition $(\cR,+)$ and a monoid under multiplication $(\cR,\cdot)$, {\it i.e.} there is not necessarily a multiplicative inverse. Additionally, multiplication is distributive under addition. The rings we consider in this paper will all be commutative under multiplication.

An ideal $\cI$ is any subset of a ring which along with the addition operation $(\cI,+)$ forms a subgroup of $(\cR,+)$ and satisfies
\begin{equation}
\forall x \in \cI, \forall y \in \cR: x \cdot y \in \cI  \text{ and } y \cdot x \in \cI.
\end{equation}
The ideal generated by a set of elements $\{ X_i \}$, is denoted $\langle X_i\rangle$ and is the minimal ideal containing the elements $X_i$; more precisely, $\langle X_i \rangle$ is the intersection of all ideals containing $\{X_i\}$. In other words the elements of an ideal generated by $\{ X_i \}$ are $\sum_i a_i X_i$ for all possible $a_i \in \cR$.

For any ring $\cR$ and ideal $\cI$, the quotient ring $\cR / \cI$ is the ring $\cR$ modulo an equivalence relation which identifies two elements if their difference is an element of $\cI$. As a simple example, consider $\bC^2$ with coordinates $(x,y)$. The space of holomorphic polynomials on $\bC^2$ with complex coefficients corresponds to the ring of polynomials in two variables with complex coefficients, $\bC [x,y]$\footnote{Throughout this paper when we refer to the ring of holomorphic polynomials on a space we mean the space of holomorphic polynomials on this space with multiplication and addition defined in the usual way. Also, we will discuss the generating function for the ring of holomorphic polynomials on a space. This generating function will have one term for each basis holomorphic monomial in the ring.}. The space of holomorphic polynomials on $\bC^2$ has as a linear basis of monomials of the form $x^my^n$, with $m,n \in \bZ_{\geq 0}$. One ideal of $\bC [x,y]$ is the ideal generated by $y$, $\cI = \langle y \rangle$. This ideal contains $y$ and anything with a factor of $y$ in it. The quotient ring $\bC[x,y] / \cI$ has a linear basis monomials of the form $x^m$, with $m \in \bZ_{\geq 0}$. In other words, $\bC[x,y] /\langle y \rangle \cong \bC [ x ]$.
%%%%%%%%%%%%%%%%%%%%%%%%%%%%%%%%%%%%%%%%%%%%%%%%%%%%%%%%%%%%%%%%%%%
\subsection{Chiral Ring}
\label{subsec:revcr}
%%%%%%%%%%%%%%%%%%%%%%%%%%%%%%%%%%%%%%%%%%%%%%%%%%%%%%%%%%%%%%%%%%%
%%%%%%%%%%%%%%%%%%%%%%%%%%%%%%%%%%%%%%%%%%%%%%%%%%%%%%%%%%%%%%%%%%%
We now review some basic facts about chiral rings. For reviews see \cite{Cachazo:2002ry, Argurio:2003ym}.

A chiral operator is any operator that is annihilated by the supersymmetry generators of one chirality, $\overline{Q}_{\dot{\alpha}}$. The OPE of chiral operators is non-singular and thus we can define a ring of chiral operators with a multiplication operation. Since an OPE of chiral operators does not depend on the positions of the operators, cluster decomposition implies that the OPE only depends on the vevs of fields. Thus, within the chiral ring, operators with the same vev are considered equivalent. As a consequence of the vacuum being annihilated by supersymmetry generators, chiral operators should be considered equivalent if they differ by a term of the form $\{\overline{Q}_{\dot{\alpha}},\dots]$; two operators $\cO_1$ and $\cO_2$ are equivalent if $\cO_1 = \cO_2 + \{\overline{Q}_{\dot{\alpha}}, X^{\dot{\alpha}} ]$.

In superspace, the condition that a superfield $\Phi$ is a chiral superfield is $\overline{D}_{\dot{\alpha}}\Phi=0$. Two chiral operators being equivalent if they differ by $\{\overline{Q}_{\dot{\alpha}},...]$ implies that the two chiral superfields $X_1$, $X_2$ they belong to are equivalent if $X_1 = X_2 + \overline{D}_{\dot{\alpha}} \overline{D}^{\dot{\alpha}} Z$. In Wess-Zumino models the equation of motion of a chiral superfield $\Phi$ is
\begin{equation}
\partial _{\Phi} W(\Phi) = \overline{D}_{\dot{\alpha}} \overline{D}^{\dot{\alpha}} \overline{ \Phi},
\end{equation}
from which we can see that in the chiral ring the F-term relations
\begin{equation}
\partial _{\Phi} W(\Phi) = 0
\end{equation}
are satisfied. If we take the gauge-variant F-terms and contract with all possible operators that result in gauge-invariant terms then we can define the ideal, $\cI_0$, which is generated by these gauge-invariant terms. Then if $\cR_0$ is the ring of chiral gauge-invariant operators, $\cR = \cR_0 / \cI_0$ is the chiral ring of a theory with nonzero superpotential. For a theory with no superpotential there are no F-terms and so the chiral ring is $\cR = \cR_0$.

For the theories we study in this paper, all elements of the chiral ring will be either single- or multi-trace operators, {\it i.e.,} a single trace of products of operators or several single-trace operators multiplied by each other. We will not consider determinants because the theories in questions will have unitary, not special unitary, gauge groups.
%%%%%%%%%%%%%%%%%%%%%%%%%%%%%%%%%%%%%%%%%%%%%%%%%%%%%%%%%%%%%%%%%%%
\subsection{Generating Functions and Plethystics}
\label{subsec:revgf}
%%%%%%%%%%%%%%%%%%%%%%%%%%%%%%%%%%%%%%%%%%%%%%%%%%%%%%%%%%%%%%%%%%%
%%%%%%%%%%%%%%%%%%%%%%%%%%%%%%%%%%%%%%%%%%%%%%%%%%%%%%%%%%%%%%%%%%%
In \cite{Benvenuti:2006qr, Feng:2007ur} the authors describe a method of counting operators in a theory using generating functions. Such a generating function typically looks like
\begin{equation}
f(t_1, \dots, t_k) = \sum_{i_1,\dots, i_k} c_{i_1, \dots, i_k} t_{1}^{i_1} \dots t_{k}^{i_k},
\end{equation}
where $t_i$ is the fugacity (chemical potential) for the $i$-th quantum number and $c_{i_1, \dots i_k}$ gives the number of operators with quantum numbers $(i_1, \dots, i_k )$.

A useful tool in  \cite{Benvenuti:2006qr} is the  ``plethystic exponential'', which is used to get the generating function for multi-trace operators from the generating function for single-trace operators at large $N$.  If we have some function $F_S(t_i)$, the plethystic exponential of the function is defined to be
\begin{equation}
\pe \left[F_S(t_i)\right] = \exp \left\{ \sum_{k=1}^{\infty} \frac{F_S(t_i^k) - F_S(0)}{k}  \right\}.
\end{equation}
To see how this gives the multi-trace operator generating function from the single-trace operator generating function, consider a generating function for single-trace operators where each single-trace operator has a different chemical potential. In this case, $F_S (t_i) = \sum_{i} t_i$. The plethystic exponential is thus
\begin{equation}
F_M(t_i) = \pe \left[F_S(t_i)\right] = \prod_{i} \frac{1}{1-t_i}.
\end{equation}
This will give one term for each way the $t_i$ can be raised to different powers and multiplied, so this is indeed the generating function for multi-trace operators.

In \cite{Benvenuti:2006qr}, the authors identify  the set of large $N$ single-trace operators with  the set of holomorphic polynomials on the moduli space, which in turn is identified with   the set of holomorphic polynomials on the transverse space. They use this logic to derive the generating function for single-trace operators using results from algebraic geometry.  We will find that these relations between 
moduli space, transverse geometry and single traces are only true modulo subtleties due to the existence of multiple branches of moduli space which we will describe.

%%%%%%%%%%%%%%%%%%%%%%%%%%%%%%%%%%%%%%%%%%%%%%%%%%%%%%%%%%%%%%%%%%%
\section{\texorpdfstring{$\cN=4$}{N=4} SYM}
\label{sec:n=4}
%%%%%%%%%%%%%%%%%%%%%%%%%%%%%%%%%%%%%%%%%%%%%%%%%%%%%%%%%%%%%%%%%%%
%%%%%%%%%%%%%%%%%%%%%%%%%%%%%%%%%%%%%%%%%%%%%%%%%%%%%%%%%%%%%%%%%%%
We begin with $U(N)$ $\cN=4$ SUSY Yang-Mills (SYM) as a particularly simple example which will illustrate the ideas used throughout the remainder of the paper. This is the worldvolume theory on a flat stack of $N$ D3-branes, with transverse space $\bC^3$. 
\begin{figure}[ht]
\centering
\includegraphics[scale=0.5]{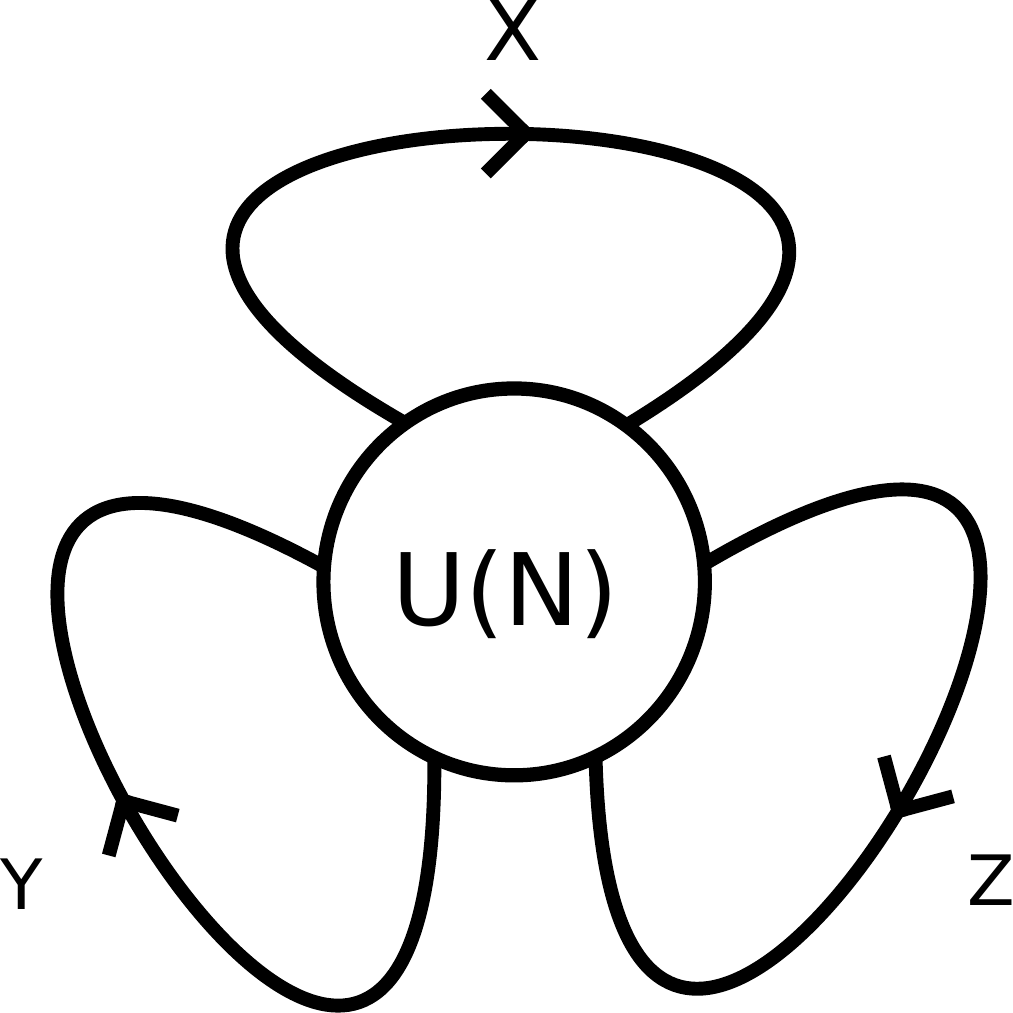}
\caption{The $\cN=1$ quiver diagram for $\mathcal{N}=4$ SYM.}
\label{fig:quivc3}
\end{figure}
In $\cN=1$ language, we can write the theory as a $U(N)$ gauge theory with three adjoint chiral superfields $X,Y,Z$ whose quiver diagram is given in figure \ref{fig:quivc3}. The $\cN=1$ superpotential is
\begin{equation}
W=\Tr \left( X \left[ Y,Z \right] \right),
\end{equation}
which yields the F-term equations 
\begin{align}
XY=YX, && XZ = ZX, && YZ = ZY,
\end{align}
where we have suppressed gauge indices.

%%%%%%%%%%%%%%%%%%%%%%%%%%%%%%%%%%%%%%%%%%%%%%%%%%%%%%%%%%%%%%%%%%%
\subsection{\texorpdfstring{$N=1$}{N=1} Moduli Space}
\label{subsec:n=4modsp}
%%%%%%%%%%%%%%%%%%%%%%%%%%%%%%%%%%%%%%%%%%%%%%%%%%%%%%%%%%%%%%%%%%%
%%%%%%%%%%%%%%%%%%%%%%%%%%%%%%%%%%%%%%%%%%%%%%%%%%%%%%%%%%%%%%%%%%%
The F-terms enforce that the matrices $X$, $Y$ and $Z$ all commute. This means that they can be simultaneously put in to upper triangular form by a unitary transformation. The D-term constraint
\begin{equation}
[X, X^{\dagger}] + [Y,Y^{\dagger}] + [Z, Z^{\dagger}]=0
\end{equation}
then enforces that $X$, $Y$ and $Z$ must be diagonal. After diagonalising, there is still a residual $S_N$ gauge symmetry which interchanges the eigenvalues, so the moduli space is $(\mathbb{C}^3)^N/S_N$ or $\text{Sym}^N (\mathbb{C}^3)$.

For a single brane, $N=1$, so the moduli space is just $\bC ^3$. This demonstrates the first relationship we would like to highlight: the $N=1$ moduli space of the D3-brane worldvolume theory is the transverse space. For multiple branes, one can interpret the $S_N$ action as swapping the positions of the $N$ D3-branes in the transverse space $\bC^3$, with the same result.

%%%%%%%%%%%%%%%%%%%%%%%%%%%%%%%%%%%%%%%%%%%%%%%%%%%%%%%%%%%%%%%
\subsection{\texorpdfstring{$W=0$}{W=0} Large \texorpdfstring{$N$}{N} Chiral Ring}
\label{subsec:n=4w=0cr}
%%%%%%%%%%%%%%%%%%%%%%%%%%%%%%%%%%%%%%%%%%%%%%%%%%%%%%%%%%%%%%%
%%%%%%%%%%%%%%%%%%%%%%%%%%%%%%%%%%%%%%%%%%%%%%%%%%%%%%%%%%%%%%%
It is interesting to consider this theory when the superpotential is turned off but the gauge coupling remains nonzero; this breaks $\cN=4$ SUSY but preserves $\cN=1$. In this situation, we now look to find the generating function for multi-trace operators. Now, the operators $X,Y,Z$ do not commute. Thus, the single-trace operators in the theory consist of various configurations of the adjoint chiral superfields with given orderings. The generating function for single-trace-operators in the large $N$ theory is then given by the generating function for 3-ary necklaces of beads\footnote{A necklace is an arrangement of objects (or beads) that is invariant under the action of the cyclic group. The generating function for $k$-ary necklaces of beads counts how many necklaces we can construct using beads of $k$ different colours.}, a problem whose solution can be found in combinatorics. The generating function is found using the P\'{o}lya enumeration theorem \cite{Redfield:1927, Polya:1937}\footnote{An introductory treatment of the theorem along with a wide range of applications can be found in \cite{Polya:1987}.}, but before describing this theorem we first introduce a few ideas.

For a finite group $G \subseteq S_n$ the cycle index is defined as
\begin{equation}
Z_G (t_1,t_2,...,t_n) \equiv \frac{1}{|G|}\sum_{g \in G} t_1^{j_1(g)}t_2^{j_2(g)}...t_n^{j_n(g)},
\end{equation}
where $j_i(g)$ is the number of cycles of length $i$ in $g$. Let $X$ be a set of $n$ objects and let $G$ be a finite group that acts on $X$. Additionally, let $Y = \{c_1,...,c_k\}$ be a set of $|Y|=k$ colours so that $Y^X$ is the set of coloured arrangements of these $n$ objects. The colour generating function is defined to be $f(c_1,...,c_k)\equiv \sum_{i=1}^kc_i$. Then the P\'{o}lya enumeration theorem counts the number of orbits under $G$ of the coloured arrangements of $n$ beads. According to the theorem, the counting is given by the generating function
\begin{equation}
F_G\left( c_1,...,c_k \right) = Z_G \left( f(c_i), f(c^2_i),...,f(c^n_i) \right).
\end{equation}

We are interested in counting $k$-ary necklaces of $n$ beads so for this case the finite group is the cyclic group $G=C_n$. The cyclic group $C_n$ has $\varphi(d)$ elements of order $d$ for each divisor $d$ of $n$, where $\varphi(d)$ is the Euler totient function\footnote{The Euler totient function $\varphi (n)$ counts the number of positive integers less than or equal to $n$ that are co-prime to $n$.}. Thus the cycle index is
\begin{equation}
Z_{C_n} \left( t_1,...,t_n \right) = \frac{1}{n}\sum_{d|n} \varphi (d) \left(t_d\right)^{\frac{n}{d}}.
\end{equation}
This means that the generating function for $k$-ary necklaces of $n$ beads is given by
\begin{equation}
F_{C_n}\left(c_1,...,c_k\right) = \frac{1}{n} \sum_{d|n} \varphi(d) \left(\Sigma_{i=1}^k c_i^d\right)^{\frac{n}{d}},
\end{equation}
and the generating function for $k$-ary necklaces of any number of beads is
\begin{equation}
F_{C}\left(c_1,...,c_k\right) = \sum_{n=1}^{\infty} \frac{1}{n} \sum_{d|n} \varphi(d) \left(\Sigma_{i=1}^k c_i^d\right)^{\frac{n}{d}} = -\sum_{d=1}^{\infty}\frac{\varphi(d)}{d}\log\left[1-\left(\Sigma_{i=1}^kc_{i}^d\right)\right].
\end{equation}

For the case of $\cN=4$ SYM with $W=0$, the generating function for single-trace operators is exactly the generating function for $3$-ary necklaces of beads ({\it i.e.},~set $k=3$ in the previous formula):
\begin{equation}
\boxed{F^{(\infty)}_S\left(x,y,z\right) = - \sum_{d=1}^{\infty}\frac{\varphi(d)}{d}\log\left[ 1-\left( x^d + y^d + z^d \right) \right]}.
\label{eq:FSn=40}
\end{equation}
We can then get the generating function for multi-trace operators by taking the plethystic exponential of this function
\begin{equation}
F^{(\infty)}_M \left( x, y, z \right) = \text{P.E.} \left[ F^{(\infty)}_S\left(x,y,z\right) \right] = \exp \left\{ \sum_{k=1}^{\infty}\frac{1}{k}F^{(\infty)}_S\left(x^k,y^k,z^k\right)\right\},
\end{equation}
so that we get
\begin{equation}
\boxed{F^{(\infty)}_M \left( x, y, z \right) = \prod_{n=1}^{\infty}\frac{1}{1-\left(x^n + y^n + z^n\right)}}.
\end{equation}
These results were originally found in \cite{Bianchi:2006ti}.

We can get back to the single-trace operator generating function by using the plethystic logarithm:
\begin{equation}
F^{(\infty)}_S(x,y,z) = PE^{-1}\left[F^{(\infty)}_M(x,y,z)\right] = \sum_{k=1}^{\infty} \frac{\mu (k)}{k} \log \left( F^{(\infty)}_M(x^k,y^k,z^k)\right)
\end{equation}
and using the identity
\begin{equation}
\frac{\varphi(n)}{n} = \sum_{d|n} \frac{\mu(d)}{d}.
\end{equation}

%%%%%%%%%%%%%%%%%%%%%%%%%%%%%%%%%%%%%%%%%%%%%%%%%%%%%%%%%%%%%%%
\subsection{\texorpdfstring{$W \neq 0$}{W neq 0} Large \texorpdfstring{$N$}{N} Chiral Ring}
\label{subsec:n=4wneq0cr}
%%%%%%%%%%%%%%%%%%%%%%%%%%%%%%%%%%%%%%%%%%%%%%%%%%%%%%%%%%%%%%%
%%%%%%%%%%%%%%%%%%%%%%%%%%%%%%%%%%%%%%%%%%%%%%%%%%%%%%%%%%%%%%%
We now look to find the generating function for multi-trace operators in the theory with nonzero superpotential. Turning on the superpotential enforces the commutativity of the adjoint chiral superfields. The generating function for single-trace operators with zero superpotential in equation \eqref{eq:FSn=40} is of the form
\begin{equation}
F^{(\infty)}_S \left(x,y,z\right) = \sum_{n_1=0}^{\infty} \sum_{n_2=0}^{\infty} \sum_{n_3=0}^{\infty} c_{n_1,n_2,n_3} x^{n_1} y^{n_2} z^{n_3},
\end{equation}
which counts the number $c_{n_1,n_2,n_3}$ of single-trace operators that we can make with $n_1$ $X$'s, $n_2$ $Y$'s and $n_3$ $Z$'s. When we enforce the commutativity of operators, all these coefficients are 1. Thus the single-trace operator generating function for large $N$ $\cN=4$ SYM with nonzero superpotential is
\begin{equation}
\boxed{F^{(\infty)}_S \left(x,y,z\right) = \sum_{n_1=0}^{\infty} \sum_{n_2=0}^{\infty} \sum_{n_3=0}^{\infty} x^{n_1} y^{n_2} z^{n_3}=\frac{1}{1-x}\frac{1}{1-y}\frac{1}{1-z}}.
\label{eq:FSn=4W}
\end{equation}
This generating function is equal to the generating function for the ring of holomorphic polynomials on $\bC^3$. We also could have obtained this formula using the P\'{o}lya enumeration theorem with $G=S_n$; see Appendix \ref{app:simp}.

We can once again take the plethystic exponential of equation \eqref{eq:FSn=4W} to give us the generating function for multi-trace operators in large $N$ $\cN=4$ SYM with nonzero superpotential:
\begin{equation}
\boxed{F^{(\infty)}_M \left(x,y,z\right) = \prod_{n=1}^{\infty} \prod_{n_1=0}^{n} \prod_{n_2=0}^{n-n_1} \frac{1}{1-x^{n_1} y^{n_2} z^{n-n_1-n_2}}}.
\label{eq:FMn=4}
\end{equation}
This formula could have alternately been derived from first principles by using $F^{(\infty)}_M = \prod_{i}\left(1-t_i \right)^{-1}$, where the product is over all single-trace operators.
%%%%%%%%%%%%%%%%%%%%%%%%%%%%%%%%%%%%%%%%%%%%%%%%%%%%%%%%%%%%%%%
\subsubsection{\texorpdfstring{$U(\infty)$}{U(infty)} Fock Space}
\label{subsec:n=4fock}
%%%%%%%%%%%%%%%%%%%%%%%%%%%%%%%%%%%%%%%%%%%%%%%%%%%%%%%%%%%%%%%
%%%%%%%%%%%%%%%%%%%%%%%%%%%%%%%%%%%%%%%%%%%%%%%%%%%%%%%%%%%%%%%
In this example, the generating function for the chiral ring of the $U(1)$ theory tells us the operator content of the theory; for $\cN=4$ SYM this generating function is the generating function for holomorphic polynomials on $\mathbb{C}^3$. This generating function is also the generating function for the Hilbert space of a single boson on $\mathbb{C}^3$ and gives a basis of wavefunctions for a particle on $\mathbb{C}^3$. In the spirit of \cite{Banks:1996vh}, we can interpret this Hilbert space in terms of wavefunctions of a single brane moving on the transverse space.

Equation \eqref{eq:FMn=4} tells us that the generating function for the large $N$ chiral ring is equal to the generating function for the multi-particle Fock space for bosons on $\mathbb{C}^3$ which is the Fock space of the multiple branes moving on the transverse space. More explicitly, the space of wavefunctions for the $i$-th boson on $\bC^3$ is spanned by
\begin{equation}
\psi_{i,p,q,r} (x,y,z) = x_{(i)}^p y_{(i)}^q z_{(i)}^r.
\end{equation}
The space of wavefunctions for two bosons is spanned by the symmetric sum
\begin{equation}
\frac{1}{2} \left( x_{(1)}^{p_1} y_{(1)}^{q_1} z_{(1)}^{r_1} x_{(2)}^{p_2} y_{(2)}^{q_2} z_{(2)}^{r_2} + x_{(2)}^{p_1} y_{(2)}^{q_1} z_{(2)}^{r_1} x_{(1)}^{p_2} y_{(1)}^{q_2} z_{(1)}^{r_2} \right).
\end{equation}
This space has a one-to-one correspondence with the Hilbert space of two bosons on $\bC^3$, which is spanned by
\begin{equation}
B_{p_1,q_1,r_1} ^{\dagger} B_{p_2,q_2,r_2} ^{\dagger} \ket{0},
\end{equation}
where $B^\dagger_{p,q,r}$ is the creation operator for a particle with wavefunction $x^p y^q z^r$, satisfying $[B_{p_1,q_1,r_1} ^{\dagger}, B_{p_2,q_2,r_2} ^{\dagger}] = 0$.

More generally, for $n$ bosons the symmetrised wavefunctions
\begin{equation}
\frac{1}{n!} \left( \sum_{\sigma \in S_n} \prod_{i=1}^n x_{\sigma(i)}^{p_i} y_{\sigma(i)}^{q_i} z_{\sigma(i)}^{r_i} \right)
\end{equation}
are in one-to-one correspondence with the Hilbert space of $n$ bosons on $\bC^3$ spanned by
\begin{equation}
B_{p_1,q_1,r_1} ^{\dagger} \cdots B_{p_n,q_n,r_n} ^{\dagger} \ket{0}.
\end{equation}
By inspecting equation \eqref{eq:FMn=4} one can see that 
\begin{equation}
F^{(\infty)}_M (x,y,z) = F_{\text{Fock}} (\bC^3).
\end{equation}
where $F_{\text{Fock}}$ is the generating function for the Fock space of bosons on $\bC^3$.
%%%%%%%%%%%%%%%%%%%%%%%%%%%%%%%%%%%%%%%%%%%%%%%%%%%%%%%%%%%%%%%
\subsection{\texorpdfstring{$N=1$}{N=1} Chiral Ring}
\label{subsec:n=4u1cr}
%%%%%%%%%%%%%%%%%%%%%%%%%%%%%%%%%%%%%%%%%%%%%%%%%%%%%%%%%%%%%%%
%%%%%%%%%%%%%%%%%%%%%%%%%%%%%%%%%%%%%%%%%%%%%%%%%%%%%%%%%%%%%%%

The generating function for operators in the chiral ring of the $U(1)$ $\cN=4$ SYM theory is simply equal to the large $N$ single-trace operator generating function given in equation \eqref{eq:FSn=4W}. This is because there is a mapping which maps every operator in the $U(1)$ $\cN=4$ SYM theory to a single-trace operator in the $U(\infty)$ theory\footnote{We use the notation $U(\infty)$ to denote the large $N$ theory.}. This mapping is
\begin{equation}
X^{n_1}Y^{n_2}Z^{n_3} \rightarrow \Tr ( X^{n_1}Y^{n_2}Z^{n_3} ).
\end{equation}
Although this mapping is rather intuitive here,  we will see in later sections that na\"ive intuition fails for more complicated theories, and in fact the generating function for the chiral ring of the $U(1)$ theory is not equal to the large $N$ single-trace operator generating function.

This highlights two more relationships that are part of this story. The first is that the $U(1)$ chiral ring is equal to the ring of holomorphic polynomials on the transverse space. The second is that the set of elements in the $U(1)$ chiral ring is equal to the set of single-trace operators in the large $N$ gauge theory.

%%%%%%%%%%%%%%%%%%%%%%%%%%%%%%%%%%%%%%%%%%%%%%%%%%%%%%%%%%%%%%%
\subsection{Conclusion}
\label{subsec:n=4con}
%%%%%%%%%%%%%%%%%%%%%%%%%%%%%%%%%%%%%%%%%%%%%%%%%%%%%%%%%%%%%%%
%%%%%%%%%%%%%%%%%%%%%%%%%%%%%%%%%%%%%%%%%%%%%%%%%%%%%%%%%%%%%%%
In this section, we observed the following relationships for $U(N)$ $\cN=4$ SYM:
\begin{enumerate}
\item The $N=1$ moduli space is the same as the space transverse to the D3-branes.
\item The $U(1)$ chiral ring is equal to the ring of holomorphic polynomials on the moduli space, and thus equal to the ring of holomorphic polynomials on the transverse space.
\item The set of elements in the $U(1)$ chiral ring is equal to the set of single-trace operators in the $U(\infty)$ theory.
\item The multi-trace operator generating function for the $U(\infty)$ theory gives us a generating function for bosons moving on the transverse space.
\end{enumerate}
In the coming sections we see while some of these relationships persist in more complicated examples, some of them do not.

%%%%%%%%%%%%%%%%%%%%%%%%%%%%%%%%%%%%%%%%%%%%%%%%%%%%%%%%%%%%%%%%%%%
\section{Conifold}
\label{sec:con}
\begin{figure}[ht]
\centering
\includegraphics[scale=0.5]{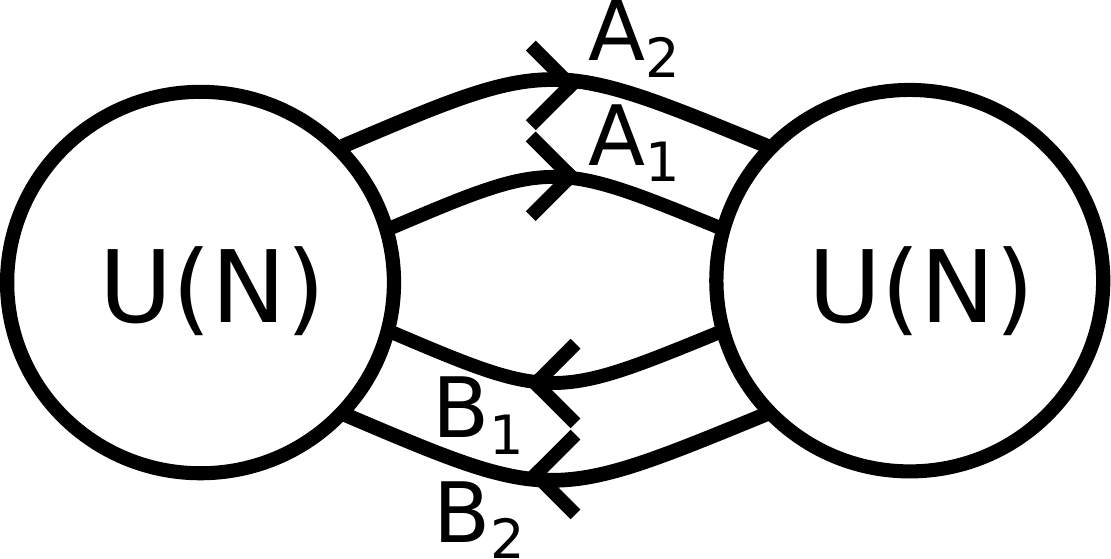}
\caption{$\cN=1$ quiver diagram for the conifold theory.}
\label{fig:quivcon}
\end{figure}

%%%%%%%%%%%%%%%%%%%%%%%%%%%%%%%%%%%%%%%%%%%%%%%%%%%%%%%%%%%%%%%%%%%
%%%%%%%%%%%%%%%%%%%%%%%%%%%%%%%%%%%%%%%%%%%%%%%%%%%%%%%%%%%%%%%%%%%
Our next example is a stack of $N$ D3-branes transverse to the conifold $\cC$, as studied in \cite{Klebanov:1998hh}.
The worldvolume theory is an $\cN=1$ $U(N) \times U(N)$ gauge theory with two chiral superfields, $A_1$ and $A_2$, in the $(\mathbf{N},\overline{\mathbf{N}})$ representation and two chiral superfields, $B_1$ and $B_2$, in the $(\overline{\mathbf{N}},\mathbf{N})$ representation. The quiver for this theory is given in figure \ref{fig:quivcon}.

The theory also has the superpotential
\begin{equation}
W = \Tr \left( A_1B_1A_2B_2 - A_1B_2A_2B_1 \right),
\end{equation}
so the F-term relations are
\begin{align}
B_1A_2B_2 - B_2A_2B_1 = 0, && B_1A_1B_2 - B_2A_1B_1 = 0, \nonumber \\
A_1B_2A_2 - A_2B_2A_1 = 0, && A_1B_1A_2 - A_2B_1A_1 = 0.
\end{align}
Gauge-invariant operators must consist of combinations of 
\begin{align}
W = A_1B_1, && X = A_1B_2, \nonumber \\
Y = A_2B_1, && Z = A_2B_2,
\label{eq:wxyzab}
\end{align}
which are in the $\mathbf{N} \otimes \overline{\mathbf{N}}$ representation of the one of the $U(N)$ gauge groups, although we have suppressed the indices.

The F-term relations expressed in terms of these are
\begin{equation}
[W,X]=[W,Y]=[W,Z]=[X,Y]=[X,Z]=[Y,Z]=0
\end{equation}
and
\begin{equation}
WZ=XY.
\end{equation}
%%%%%%%%%%%%%%%%%%%%%%%%%%%%%%%%%%%%%%%%%%%%%%%%%%%%%%%%%%%%%%%%%%%
\subsection{\texorpdfstring{$N=1$}{N=1} Moduli Space}
\label{subsec:conmodsp}
%%%%%%%%%%%%%%%%%%%%%%%%%%%%%%%%%%%%%%%%%%%%%%%%%%%%%%%%%%%%%%%%%%%
%%%%%%%%%%%%%%%%%%%%%%%%%%%%%%%%%%%%%%%%%%%%%%%%%%%%%%%%%%%%%%%%%%%
In the $N=1$ ({\it i.e.},~$U(1)^2$) theory, the superpotential vanishes. Thus we need only solve the D-term equation
\begin{equation}
D= |A_1|^2 + |A_2|^2 - |B_1|^2 - |B_2|^2=0,
\label{eq:conD}
\end{equation}
which is the equation for the conifold. (We do not consider an FI term.) We once again see that the $N=1$ moduli space is exactly the same as the space transverse to the brane.
%%%%%%%%%%%%%%%%%%%%%%%%%%%%%%%%%%%%%%%%%%%%%%%%%%%%%%%%%%%%%%%
\subsection{\texorpdfstring{$W=0$}{W=0} Large \texorpdfstring{$N$}{N} Chiral Ring}
\label{subsec:conw=0cr}
%%%%%%%%%%%%%%%%%%%%%%%%%%%%%%%%%%%%%%%%%%%%%%%%%%%%%%%%%%%%%%%
%%%%%%%%%%%%%%%%%%%%%%%%%%%%%%%%%%%%%%%%%%%%%%%%%%%%%%%%%%%%%%%
For the theory with zero superpotential, the generating function for single-trace operators is given by the generating function for $4$-ary necklaces:
\begin{equation}
\boxed{F^{(\infty)}_S\left(w,x,y,z\right) = - \sum_{d=1}^{\infty}\frac{\varphi(d)}{d}\log\left[ 1-\left( w^d + x^d + y^d + z^d \right) \right]}.
\label{eq:confs}
\end{equation}
Rewriting $w,x,y,$ and $z$ in terms of $a_{1,2}$ and $b_{1,2}$ via the relations in \eqref{eq:wxyzab}, the generating function is of the form
\begin{equation}
F^{(\infty)}_S\left(a_1,a_2,b_1,b_2\right) = \sum_{n=0}^{\infty} \sum_{n_1=0}^{n} \sum_{n_2=0}^{n} c_{n, n_1, n_2}, a_1^{n_1}b_1^{n_2}a_2^{n-n_1}b_2^{n-n_2},
\label{eq:fscongen}
\end{equation}
where $c_{n, n_1, n_2}$ counts the number of single-trace operators that can be constructed using $n_1$ $A_1$ operators, $n-n_1$ $A_2$ operators, $n_2$ $B_1$ operators, $n-n_2$ $B_2$ operators. The number of $A$ operators and the number of $B$ operators must be equal for the single-trace operator to be gauge-invariant. The generating function for multi-trace operators is obtained by taking the plethystic exponential of \eqref{eq:confs}:
\begin{equation}
\boxed{F^{(\infty)}_M \left( x, y, z \right) = \prod_{n=1}^{\infty}\frac{1}{1-\left(w^n + x^n + y^n + z^n\right)}}.
\end{equation}
This is in agreement with \cite{Pasukonis:2013ts}.
%%%%%%%%%%%%%%%%%%%%%%%%%%%%%%%%%%%%%%%%%%%%%%%%%%%%%%%%%%%%%%%
\subsection{\texorpdfstring{$W \neq 0$}{W neq 0} Large \texorpdfstring{$N$}{N} Chiral Ring}
\label{subsec:conwneq0cr}
%%%%%%%%%%%%%%%%%%%%%%%%%%%%%%%%%%%%%%%%%%%%%%%%%%%%%%%%%%%%%%%
%%%%%%%%%%%%%%%%%%%%%%%%%%%%%%%%%%%%%%%%%%%%%%%%%%%%%%%%%%%%%%%
All single-trace operators in the conifold theory have alternating $A$'s and $B$'s, {\it i.e.} they are of the form $\Tr(ABAB...AB)$. Turning on the superpotential means that the F-term relations allow us to organise the trace so that the first $n_1$ $A$ operators are $A_1$'s and the last $n-n_1$ $A$ operators are $A_2$'s, and similarly for the $B$'s. Thus the generating function is just the function in equation \eqref{eq:fscongen} with all the coefficients set to 1:
\begin{equation}
\boxed{F^{(\infty)}_S(a_1,a_2,b_1,b_2)=\sum_{n=0}^{\infty}\sum_{n_1=0}^{n}\sum_{n_2=0}^{n}a_1^{n_1}b_1^{n_2}a_2^{n-n_1}b_2^{n-n_2}}.
\label{eq:FScon}
\end{equation}
This formula has a closed form expression in terms of $w$, $x$, $y$ and $z$:
\begin{equation}
F^{(\infty)}_S(w,x,y,z)=\frac{1}{w-x-y+z}\left(\frac{1}{1-w}-\frac{1}{1-x}-\frac{1}{1-y}+\frac{1}{1-z}\right).
\label{eq:FSconrat}
\end{equation}
The generating function for multi-trace operators can once again be found by taking the plethystic exponential of equation \eqref{eq:FScon}:
\begin{equation}
\boxed{F^{(\infty)}_M(a_1,a_2,b_1,b_2)=\prod_{n=0}^{\infty} \prod_{n_1=0}^{n} \prod_{n_2=0}^{n} \frac{1}{1-a_1^{n_1}b_1^{n_2}a_2^{n-n_1}b_2^{n-n_2}}},
\label{eq:FMcon}
\end{equation}
which can again be seen intuitively from $F^{(\infty)}_M = \prod_{\Phi}\left(1-\Phi\right)^{-1}$, where the product is over single-trace operators. Also as before, $F^{(\infty)}_S$ can alternately be derived from the P\'{o}lya enumeration theorem by taking the product of two generating functions for 2-ary necklaces.

If we take equation \eqref{eq:FSconrat} and make the substitutions $w \rightarrow qa$, $x \rightarrow qb$, $y \rightarrow \frac{q}{b}$, and $z \rightarrow \frac{q}{a}$ we regain the form of the generating function presented in \cite{Benvenuti:2006qr}:
\begin{equation}
F^{(\infty)}_S(a,b,q) = \frac{a b (q-1) (q+1)}{(a-q) (a q-1) (q-b) (b q-1)}.
\end{equation}
This substitution is indicative of the relationships between the charges that we have chosen here and the charges that are chosen in \cite{Benvenuti:2006qr}.
%%%%%%%%%%%%%%%%%%%%%%%%%%%%%%%%%%%%%%%%%%%%%%%%%%%%%%%%%%%%%%%
\subsubsection{\texorpdfstring{$U(\infty)$}{U(infty)} Fock Space}
\label{subsec:confock}
%%%%%%%%%%%%%%%%%%%%%%%%%%%%%%%%%%%%%%%%%%%%%%%%%%%%%%%%%%%%%%%
%%%%%%%%%%%%%%%%%%%%%%%%%%%%%%%%%%%%%%%%%%%%%%%%%%%%%%%%%%%%%%%
Equation \eqref{eq:FMcon} tells us that the generating function for multi-trace operators in the $U(\infty)$ theory is equal to the generating function for the Fock space of bosons on the conifold:
\begin{equation}
F^{(\infty)}_M(a_1,a_2,b_1,b_2) = F^{(\infty)}_{\text{Fock}}(\cC).
\end{equation}
As was the case for $\cN=4$ SYM, we can again interpret this as the Fock space for branes on the transverse space.
%%%%%%%%%%%%%%%%%%%%%%%%%%%%%%%%%%%%%%%%%%%%%%%%%%%%%%%%%%%%%%%
\subsection{\texorpdfstring{$N=1$}{N=1} Chiral Ring}
\label{subsec:conu1cr}
%%%%%%%%%%%%%%%%%%%%%%%%%%%%%%%%%%%%%%%%%%%%%%%%%%%%%%%%%%%%%%%
%%%%%%%%%%%%%%%%%%%%%%%%%%%%%%%%%%%%%%%%%%%%%%%%%%%%%%%%%%%%%%%
As with $\cN=4$, there is a one-to-one mapping between operators in the $U(1)$ theory and single-trace operators in the $U(\infty)$ theory. This mapping is 
\begin{equation}
A_1^{n_1}A_2^{n-n_1}B_1^{n_2}B_2^{n-n_2} \rightarrow \Tr ( A_1^{n_1}A_2^{n-n_1}B_1^{n_2}B_2^{n-n_2} ).
\end{equation}
Thus the generating function for the chiral ring of the $U(1)$ theory is equal to the generating function for single-trace operators in the $U(\infty)$ theory given in equation \eqref{eq:FScon}.

%%%%%%%%%%%%%%%%%%%%%%%%%%%%%%%%%%%%%%%%%%%%%%%%%%%%%%%%%%%%%%%
\subsection{Conclusion}
\label{subsec:concon}
%%%%%%%%%%%%%%%%%%%%%%%%%%%%%%%%%%%%%%%%%%%%%%%%%%%%%%%%%%%%%%%
%%%%%%%%%%%%%%%%%%%%%%%%%%%%%%%%%%%%%%%%%%%%%%%%%%%%%%%%%%%%%%%
We see from this slightly more complicated example many of the same phenomena that we saw with $\cN=4$ SYM. First, the moduli space is equal to the transverse space, and  the chiral ring of the $U(1)$ theory is the ring of holomorphic polynomials on this space. The chiral ring has the same elements as the set of single-trace operators in the large $N$ theory. Also, in the $U(\infty)$ theory we can identify  a Fock space of bosons which we interpret as the Fock space of the branes moving on the transverse space. In the next section, we will see that some of these relationships do not hold more generally.
%%%%%%%%%%%%%%%%%%%%%%%%%%%%%%%%%%%%%%%%%%%%%%%%%%%%%%%%%%%%%%%%%%%
\section{\texorpdfstring{$\bC^3 / \bZ_2$}{C3/Z2}}
\label{sec:c3z2}
%%%%%%%%%%%%%%%%%%%%%%%%%%%%%%%%%%%%%%%%%%%%%%%%%%%%%%%%%%%%%%%%%%%
%%%%%%%%%%%%%%%%%%%%%%%%%%%%%%%%%%%%%%%%%%%%%%%%%%%%%%%%%%%%%%%%%%%

We now consider the theory living on the worldvolume of $N$ D3-branes probing a $\bC^3 / \bZ_2$ singularity \cite{Douglas:1997de}.
\begin{figure}[ht]
\centering
\includegraphics[scale=0.5]{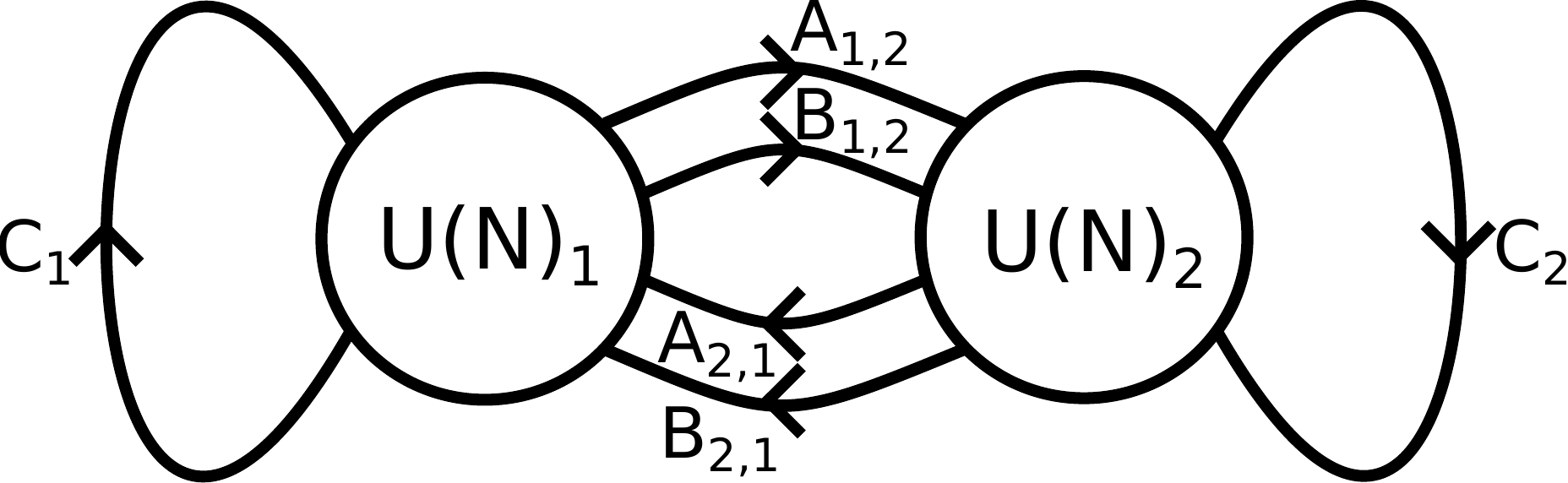}
\caption{$\cN=1$ quiver diagram for the $\mathbb{C}_3/\mathbb{Z}_2$ theory.}
\label{fig:quivc3z2}
\end{figure}
This theory has a quiver as in figure \ref{fig:quivc3z2} and a superpotential given by
\begin{equation}
W = \Tr \left[ C_1 (A_{1,2} B_{2,1} - B_{1,2} A_{2,1}) +C_2 (A_{2,1} B_{1,2} - B_{2,1} A_{1,2}) \right].
\label{eq:wc2z2}
\end{equation}
From the quiver and superpotential we can see that this theory is in fact an $\cN=2$ theory, and can flow to the conifold theory via mass terms for the adjoint chiral superfields. The F-term relations are
\begin{align}
C_1 A_{1,2} &= A_{1,2}C_2, & C_1 B_{1,2} &= B_{1,2}C_2, \nonumber \\
C_2A_{2,1} &= A_{2,1}C_1, & C_2B_{2,1} &= B_{2,1}C_1, \nonumber \\
A_{1,2}B_{2,1} &=B_{1,2}A_{2,1}, & A_{2,1}B_{1,2} &= B_{2,1}A_{1,2}.
\label{eq:c3z3f}
\end{align}
It will prove useful to use the composite operators $W=B_{1,2}A_{2,1}$, $X=A_{1,2}A_{2,1}$, $Y=B_{1,2}B_{2,1}$ and $Z=A_{1,2}B_{2,1}$ throughout the remainder of this section.
%%%%%%%%%%%%%%%%%%%%%%%%%%%%%%%%%%%%%%%%%%%%%%%%%%%%%%%%%%%%%%%%%%%
\subsection{\texorpdfstring{$N=1$}{N=1} Moduli Space}
\label{subsec:c3z2modsp}
\begin{figure}[ht]
\centering
\includegraphics[scale=0.5]{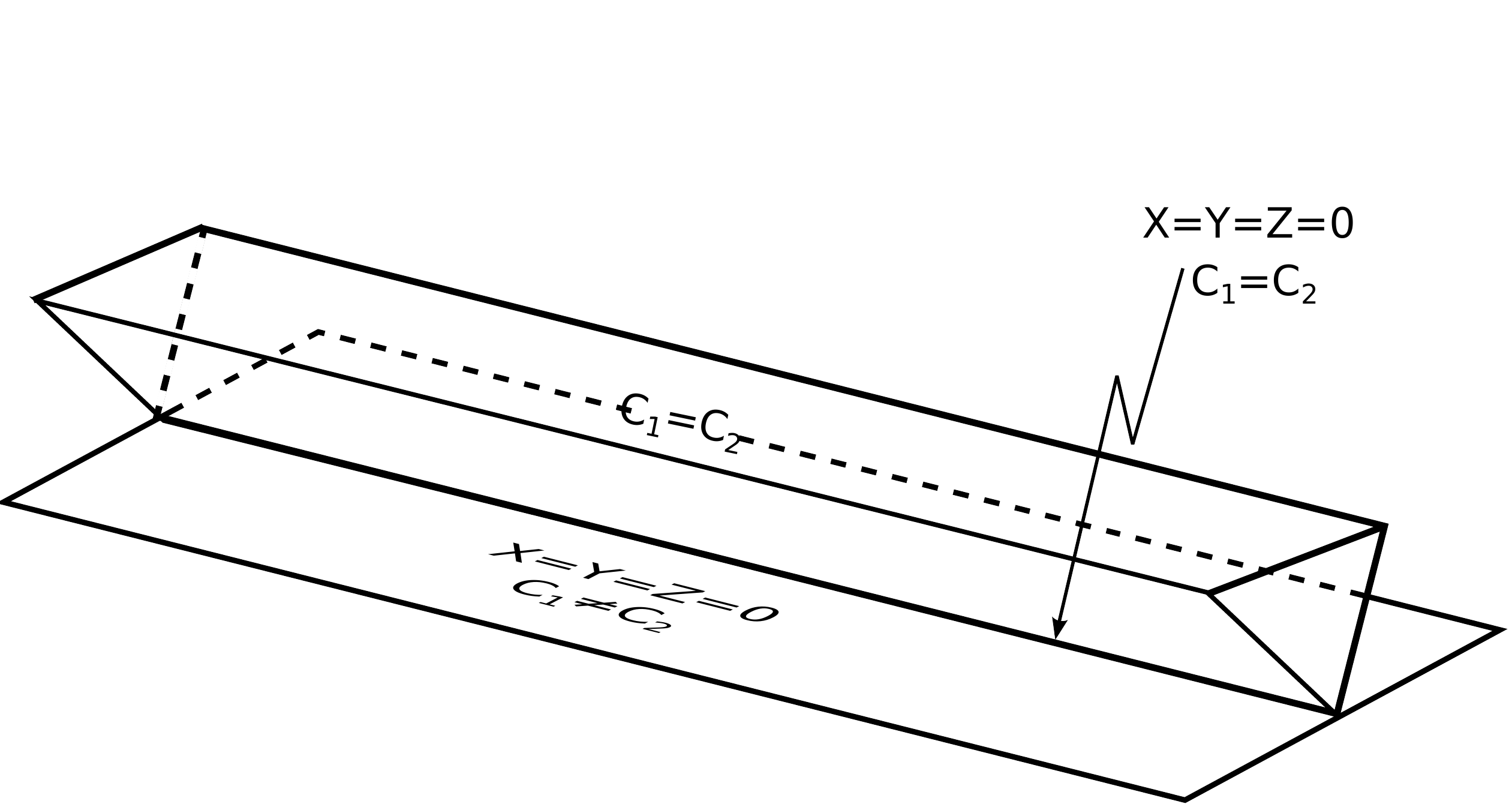}
\caption{Moduli space of the $\mathbb{C}^3/\mathbb{Z}_2$ theory. The $C_1=C_2$ branch is 3 complex dimensional and the $C_1 \neq C_2$ branch is 2 complex dimensional.}
\label{fig:c3z2mod}
\end{figure}

%%%%%%%%%%%%%%%%%%%%%%%%%%%%%%%%%%%%%%%%%%%%%%%%%%%%%%%%%%%%%%%%%%%
%%%%%%%%%%%%%%%%%%%%%%%%%%%%%%%%%%%%%%%%%%%%%%%%%%%%%%%%%%%%%%%%%%%
The F-terms equations have two branches of solutions:
\begin{enumerate}
\item $\{ X,Y,Z,C_1,C_2 \, |\, XY=Z^2, C_1=C_2 \}$,
\item $\{ X,Y,Z,C_1,C_2 \, | \,X=Y=Z=0 \}$.
\end{enumerate}
On the first branch the moduli space is described by the gauge-invariant operators $X$, $Y$, $Z$ and $C$ ($=C_1=C_2$) subject to $XY=Z^2$; this is just the space $\mathbb{C}^3 / \mathbb{Z}_2$. On the second branch, $C_1$ does not necessarily equal $C_2$, and $X=Y=Z=0$; this is the simply $\mathbb{C}^2$. The two branches intersect along the line $C_1=C_2$ when $X=Y=Z=0$. For a cartoon of the full moduli space, see figure \ref{fig:c3z2mod}.

We denote the full moduli space $\bC^3 / \bZ_2 \cup \bC^2$, where the particular union that is meant is the one where the two spaces share the line $X=Y=Z=0$, $C_1 = C_2$. In other words
\begin{align}
\cM = \bC^3 / \bZ_2 \cup \bC^2, && \bC^3 / \bZ_2 \cap \bC^2 = \bC.
\label{eq:c3z2m}
\end{align}

In contrast to the previous two examples, we see here that the moduli space is not simply the space transverse to the D3-branes. There is one \textit{main branch} which is the transverse space, but we also see the existence of an \textit{extra branch} which has a different dimension than the main branch. This extra branch of moduli space is something we will see in later examples  and has been observed, e.g. in \cite{Berenstein:2000hy, Hanany:2006uc}.
%%%%%%%%%%%%%%%%%%%%%%%%%%%%%%%%%%%%%%%%%%%%%%%%%%%%%%%%%%%%%%%
\subsection{\texorpdfstring{$W=0$}{W=0} Large \texorpdfstring{$N$}{N} Chiral Ring}
\label{subsec:c3z2w=0cr}
%%%%%%%%%%%%%%%%%%%%%%%%%%%%%%%%%%%%%%%%%%%%%%%%%%%%%%%%%%%%%%%
%%%%%%%%%%%%%%%%%%%%%%%%%%%%%%%%%%%%%%%%%%%%%%%%%%%%%%%%%%%%%%%
The $W=0$ large $N$ single-trace operator generating function can be found using the P\'{o}lya enumeration theorem, as in previous sections but this time counting 6-ary necklaces. One wrinkle is that because the fields $C_1$ and $C_2$ are not charged under the same gauge group, the fields should not be placed next to each other in a trace. In combinatorics language, we cannot place the beads of colour $c_1$ and $c_2$ next to each other in a necklace. However, this problem is easily solved by using the colour generating function
\begin{equation}
f(w,x,y,z,c_1,c_2) = w+x+y+z+c_1+c_2-c_1c_2,
\end{equation}
where the final term subtracts the contribution from necklaces with adjacent $c_1$ and $c_2$ beads ($w,x,y,z$ are as in the previous section). This yields
\begin{equation}
\boxed{F^{(\infty)}_S(w,x,y,z,c_1,c_2) = - \sum_{k=1}^{\infty}\frac{\varphi(k)}{k}\log\left[ 1-\left( w^k + x^k + y^k + z^k + c_1^k + c_2^k - c_1^kc_2^k\right) \right]},
\end{equation}
and taking the plethystic exponential gives us the multi-trace operator generating function
\begin{equation}
\boxed{F^{(\infty)}_M (w,x,y,z,c_1,c_2) = \prod_{k=1}^{\infty} \frac{1}{1-\left(w^k + x^k + y^k + z^k + c_1^k + c_2^k - c_1^kc_2^k\right)}}.
\end{equation}
This matches the formula given in \cite{Pasukonis:2013ts}.

%%%%%%%%%%%%%%%%%%%%%%%%%%%%%%%%%%%%%%%%%%%%%%%%%%%%%%%%%%%%%%%
\subsection{\texorpdfstring{$W \neq 0$}{W neq 0} Large \texorpdfstring{$N$}{N} Chiral Ring}
\label{subsec:c3z2wneq0cr}
%%%%%%%%%%%%%%%%%%%%%%%%%%%%%%%%%%%%%%%%%%%%%%%%%%%%%%%%%%%%%%%
%%%%%%%%%%%%%%%%%%%%%%%%%%%%%%%%%%%%%%%%%%%%%%%%%%%%%%%%%%%%%%%

We now turn on the superpotential in equation \eqref{eq:wc2z2}. The F-term equivalences preserve the number of $A$'s, the number of $B$'s, and the number of ($C_1$'s +$C_2$'s), so our single-trace operator generating function can have at most three chemical potentials. All operators can be arranged using the F-term relations so that they have the form $\Tr \lp C_1^n \rp$, $\Tr \lp C_2^n \rp$ or $\Tr \lp C_1^m A_{1,2}A_{2,1}A_{1,2}B_{2,1}B_{1,2}...A_{1,2}B_{2,1} \rp$, with the order of $A$'s and $B's$ irrelevant. The generating function is then
\begin{equation}
\boxed{F^{(\infty)}_S(a,b,c) = \sum_{m=0}^{\infty} c^m \sum_{\ell=0} ^{\infty} \sum_{k=0}^{2 \ell} a^{2 \ell-k} b^k + \sum_{m=1}^{\infty} c^m = \frac{1+ab}{(1-c)(1-a^2)(1-b^2)}  + \frac{c}{1-c}},
\label{eq:c3z2wf}
\end{equation}
with plethystic exponential 
\begin{equation}
\boxed{F^{(\infty)}_M \lp a,b,c \rp = \prod_{n=0}^{\infty}\prod_{\ell=0}^{\infty} \prod_{k=0}^{2 \ell} \frac{1}{1-c^m a^{2 \ell - k}b^{k}} \prod_{n=1}^{\infty} \frac{1}{1-c^n}}.
\end{equation}

Using the methods in \cite{Benvenuti:2006qr}, one can derive the single-trace operator generating function 
\begin{equation}
F^{(\infty)}_S (t) = \frac{1+t^2}{(1-t)^3(1+t)^2}.
\label{eq:c3z2han}
\end{equation}
We have included this short calculation in appendix \ref{app:hanany}.
To compare the with our answer, we make the substitution $a \rightarrow t$, $b \rightarrow t$, $c \rightarrow t$, which yields
\begin{equation}
F^{(\infty)}_S \lp t \rp = \frac{t^5-2t^3+t^2+t+1}{(1-t)^3(1+t)^2},
\label{eq:c3z2FSt}
\end{equation}
which is different than the earlier result.

It is straightforward to find the source of the discrepancy. In \cite{Benvenuti:2006qr} it was assumed that the set of single-trace operators was equal to the set of elements in the ring of holomorphic polynomials on the moduli space and thus equal to the set of elements in the ring of holomorphic polynomials on the transverse space. As we have seen for the $\bC^3 / \bZ_2$ theory, this is not quite correct. Instead we saw that the moduli space has two separate branches, only one of which is the space transverse to the D3-branes. Indeed, we find that subtracting the contribution from the $C_1 \neq C_2$ branch reproduces the previous result.

We can alternatively see the difference in the two approaches from a ring theoretic perspective. For the full worldvolume theory, the set of single-trace operators is not equal to the set of elements in the ring of holomorphic polynomials on $\bC^3 / \bZ_2$, since $\Tr (C_1^k)$ and $\Tr(C_2^k)$ are not necessarily equal. Denoting the ring of gauge-invariant operators in the $W=0$ $\mathbb{C}^3 / \mathbb{Z}_2$ theory by $R_0$, we consider the ring $R_W = \{ R_0 | \text{F-term equivalences} \}$, while the earlier work considered only $R'_W = \{ R_W | C_1=C_2 \}$, which is the ring of holomorphic polynomials on $\bC^3 / \bZ_2$.

%%%%%%%%%%%%%%%%%%%%%%%%%%%%%%%%%%%%%%%%%%%%%%%%%%%%%%%%%%%%%%%
\subsubsection{\texorpdfstring{$U(\infty)$}{U(infty)} Fock Space}
\label{subsec:c3z2fock}
%%%%%%%%%%%%%%%%%%%%%%%%%%%%%%%%%%%%%%%%%%%%%%%%%%%%%%%%%%%%%%%
%%%%%%%%%%%%%%%%%%%%%%%%%%%%%%%%%%%%%%%%%%%%%%%%%%%%%%%%%%%%%%%
For $\cN=4$ SYM and the conifold, we saw that the multi-trace operator generating function was equal to the generating function for the Fock space for bosons moving on the transverse space. In the $\bC^3 / \bZ_2$ theory, the extra branch of moduli space changes this story. Here, we have the generating function for the Fock space for bosons moving on $\bC^3 / \bZ_2$ multiplied by the generating function for bosons moving on $\bC$:
\begin{equation}
F^{(\infty)}_M(a,b,c) = F_{\text{Fock}}(\bC^3 / \bZ_2) \times F_{\text{Fock}}(\bC).
\end{equation}
This means that the multi-trace operator generating function gives the Fock space for bosons moving on $\bC^3 / \bZ_2 \coprod \bC$, where $\coprod$ indicates a disjoint union. So wavefunctions for the bosons can be any function in the space of functions on $\bC^3 / \bZ_2$ or any function on the space of functions on $\bC$, with only one identity function.
%%%%%%%%%%%%%%%%%%%%%%%%%%%%%%%%%%%%%%%%%%%%%%%%%%%%%%%%%%%%%%%
\subsection{\texorpdfstring{$N=1$}{N=1} Chiral Ring}
\label{subsec:c3z2u1cr}
%%%%%%%%%%%%%%%%%%%%%%%%%%%%%%%%%%%%%%%%%%%%%%%%%%%%%%%%%%%%%%%
%%%%%%%%%%%%%%%%%%%%%%%%%%%%%%%%%%%%%%%%%%%%%%%%%%%%%%%%%%%%%%%
In our previous examples, the generating function for the $U(1)$ chiral ring has been equal to the generating function for single-trace operators in the large $N$ gauge theory. However, this is not the case in the present example.
In the $U(\infty)$ theory, we cannot have $\Tr (C_1^{m_1} C_2^{m_2})$ operators since $C_1$ and $C_2$ transform in different gauge groups; however when $N=1$ these fields become uncharged so that the $C_1^{m_1} C_2^{m_2}$ operator is gauge-invariant. This means that we no longer have a one-to-one mapping between operators in the $U(1)$ theory and single-trace operators in the large $N$ theory. The goal of this section is to derive the $N=1$ chiral ring which includes the operators just mentioned.

We begin with the chiral ring of the $U(1)$ theory with no superpotential, which we denote $\mathcal{R}_0$. A basis for this ring is the set of gauge-invariant monomials built out of the fields $W,X,Y,Z,C_1,C_2$, $\mathcal{R}_0 = \{W^{n_1} X^{n_2} Y^{n_3} Z^{n_4} C_1^{n_5} C_2^{n_6} | n_i \geq 0\}$. To enforce the F-terms, we mod out by an ideal generated by the relevant constraints,
\begin{equation}
\mathcal{R} = \mathcal{R}_0  / \cI_0,
\label{eq:c3z2r}
\end{equation}
where 
$\cI_0 = \langle X(C_1-C_2), Y(C_1-C_2), Z(C_1-C_2), XY-Z^2, W-Z \rangle$. Thus the full chiral ring $\mathcal{R}$ is spanned by the basis
\begin{equation}
\{ X^{n_1} Y^{n_2} Z^{n_3} C_1^{n_4} \} \cup \{ C_1^{m_1} C_2^{m_2} \},
\label{eq:c3z2cr}
\end{equation}
where $n_1, n_2, n_4 \in \mathbb{Z}_{\geq 0}$, $n_3 \in \{0,1 \}$, $m_1 \in \mathbb{Z}_{\geq 0}$, and $ m_2 \in \mathbb{Z}_+$ and the rule for multiplication is
\begin{eqnarray}
(X^{n_1}Y^{n_2}Z^{n_3}C_1^{n_4}) \cdot (C_1^{m_1}C_2^{m_2}) &=& X^{n_1}Y^{n_2}Z^{n_3}C_1^{n_4+m_1+m_2}, \quad n_1+n_2+n_3 > 0 \cr
&=& C_1^{n_4+m_1}C_2^{m_2}, \quad \quad \quad \quad \qquad n_1=n_2=n_3=0.
\end{eqnarray}
From this way of expressing the basis we see that this ring is $\cR[\bC^3 / \bZ_2 \cup \bC^2]$, the ring of holomorphic polynomials on the space $\bC^3 / \bZ_2 \cup \bC^2$. The set of elements in this ring is the union of the set of functions on $\bC^3 / \bZ_2$ and the set of functions on $\bC^2$. These sets of functions have the bases $\{X^{n_1}Y^{n_2}Z^{n_3}C_1^{n_4}\}$ and $\{ C_1^{m_1}C_2^{m_2} \}$, respectively, and share the coordinate $C_1$.

The generating function for the $U(1)$ chiral ring is then
\begin{equation}
F^{(1)} \lp x,y,z,c \rp = \sum_{n_4=0}^{\infty} \sum_{n_1=0}^{\infty} \sum_{n_2=0}^{\infty} \sum_{n_3=0}^1 c^{n_4} x^{n_1} y^{n_2} z^{n_3} + \sum_{m_1=0}^{\infty}\sum_{m_2=0}^{\infty} c^{m_1+m_2} - \sum_{m=0}^{\infty}c^m.
\label{eq:c3z2FSu1}
\end{equation}
We can see from this expression that the generating function is the sum of the generating functions for holomorphic polynomials on the two branches minus the generating function for holomorphic polynomials on the intersection. This subtraction is necessary to avoid double counting of operators.

\begin{figure}[ht]
\centering
\includegraphics[scale=0.25]{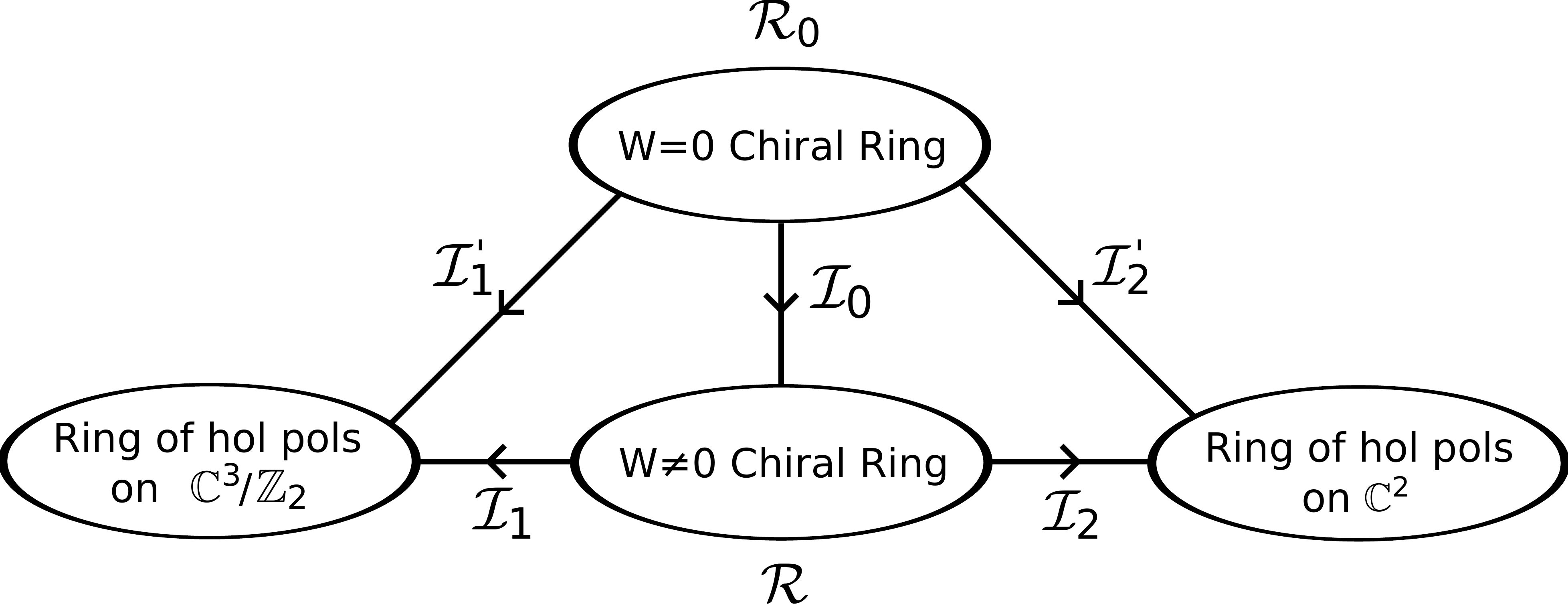}
\caption{Chiral rings for the $\bC^3 / \bZ_2$ theory. Following an arrow means quotienting by an ideal.}
\label{fig:ideals}
\end{figure}

We can further localize elements of $\mathcal{R}$ to the two branches of the moduli space by modding out by
\begin{equation}
\cI_1 = \langle C_1-C_2 \rangle = \{ C_1^{m_1} (C_1^{m_2} - C_2^{m_2}) | m_1 \in \bZ_{\geq 0}, m_2 \in \bZ_+ \}
\end{equation}
for $\bC^3 / \bZ_2$ or
\begin{equation}
\mathcal{I}_2 = \langle X,Y,Z \rangle = \{ X^{n_1}Y^{n_2}Z^{n_3}C_1^{n_4} | n_i \in \bZ_{\geq 0}, n_1+n_2+n_3 > 0 \}.
\end{equation}
for $\bC^2$. The resulting quotient rings, $\cR/\cI_1$ and $\cR / \cI_2$, are identically the rings of holomorphic functions on the two branches. An alternate description of these two quotient rings of functions utilises minimal prime ideals\footnote{An ideal $\cI$ is prime if $x \cdot y \in \cI$ implies either $x \in \cI$ or $y \in \cI$, and a prime ideal $\cI$ is a minimal prime ideal over $\cI_0$ if there does not exist another prime ideal $\cI'$ satisfying $\cI \supset \cI' \supset \cI_0$.}. Although the ideal $\cI_0$ that we originally used to quotient $\cR_0$ by is not a prime ideal, there exist two minimal prime ideals over $\cI_0$, $\cI'_1 = \langle C_1-C_2, XY-Z^2, W-Z \rangle$ and $\cI'_2 = \langle X,Y,Z,W-Z \rangle$. If we instead quotient $\cR_0$ by $\cI'_1$ or $\cI'_2$ then we would have obtained $\cR/\cI_1$ and $\cR / \cI_2$, respectively. This is illustrated in figure \ref{fig:ideals}.

%%%%%%%%%%%%%%%%%%%%%%%%%%%%%%%%%%%%%%%%%%%%%%%%%%%%%%%%%%%%%%%
\subsection{Conclusion}
\label{subsec:c3z2con}
%%%%%%%%%%%%%%%%%%%%%%%%%%%%%%%%%%%%%%%%%%%%%%%%%%%%%%%%%%%%%%%
%%%%%%%%%%%%%%%%%%%%%%%%%%%%%%%%%%%%%%%%%%%%%%%%%%%%%%%%%%%%%%%
In this section we described how some of the relationships found in previous sections no longer hold. In particular:
\begin{enumerate}
\item The moduli space has two branches. One of these branches is the transverse space $\bC^3 / \bZ_2$ and the other is $\bC^2$. This can be interpreted in terms of fractional branes as in \cite{Hanany:2006uc}.
\item The $U(1)$ chiral ring is equal to the ring of holomorphic polynomials on the moduli space and is slightly ``larger'' that the ring of holomorphic polynomials on the transverse space, in the sense that it contains the ring of holomorphic polynomials on the transverse space as a quotient ring.
\item The set of elements in the $U(1)$ chiral ring is different from the set of single-trace operators in the large $N$ theory due to the presence of adjoint fields.
\item The multi-trace operator generating function is equal to the generating function of bosons moving on the space $\text{Transverse Space} \coprod \bC$.
\end{enumerate}
We will build on the results that we have found here in the coming sections and further elucidate these connections.
%%%%%%%%%%%%%%%%%%%%%%%%%%%%%%%%%%%%%%%%%%%%%%%%%%%%%%%%%%%%%%%%%%%
\section{\texorpdfstring{$\bC^3 / \bZ_n$}{C3/Zn}}
\label{sec:c3zn}
%%%%%%%%%%%%%%%%%%%%%%%%%%%%%%%%%%%%%%%%%%%%%%%%%%%%%%%%%%%%%%%%%%%
%%%%%%%%%%%%%%%%%%%%%%%%%%%%%%%%%%%%%%%%%%%%%%%%%%%%%%%%%%%%%%%%%%%

\begin{figure}[ht]
\centering
\includegraphics[scale=0.25]{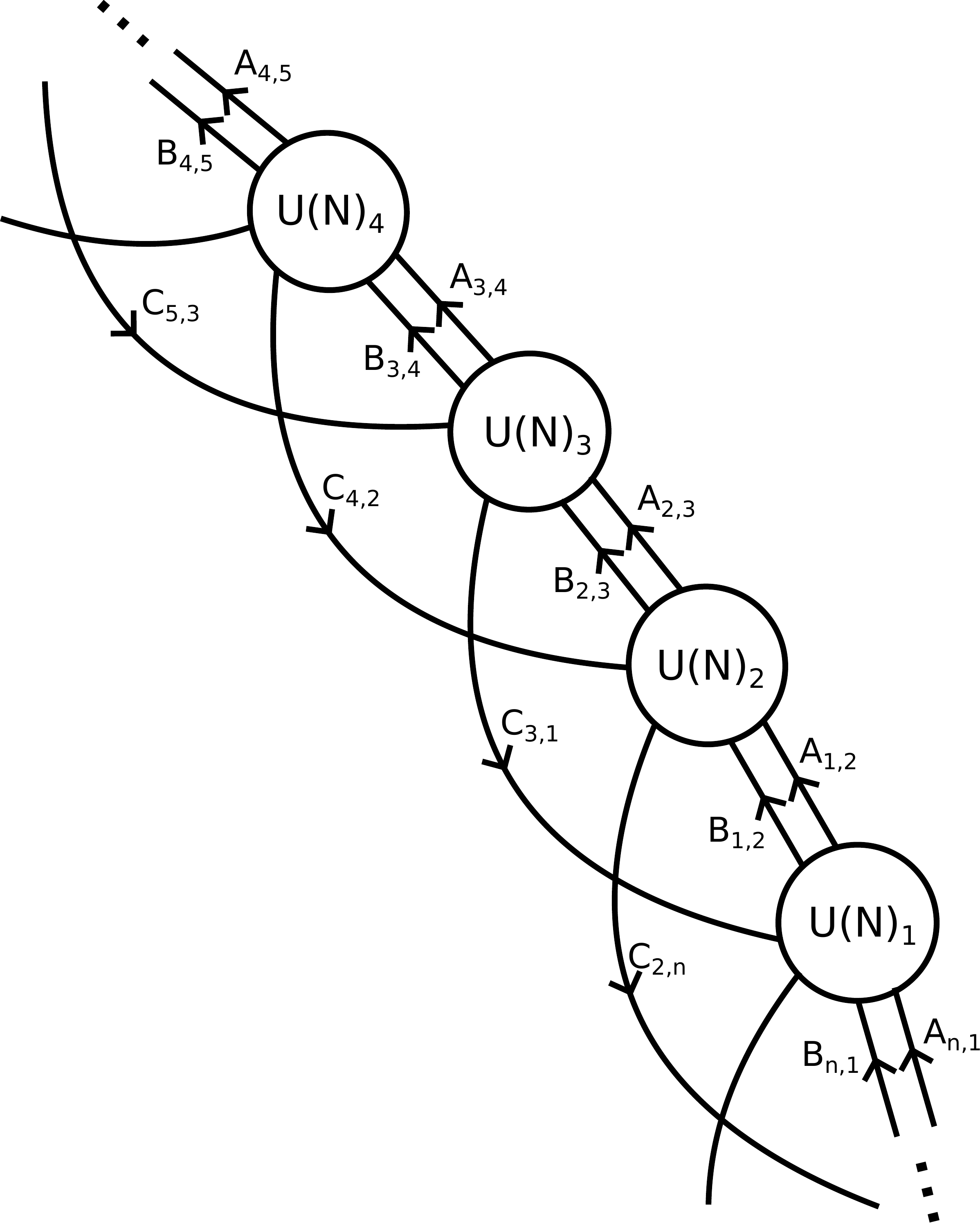}
\caption{$\cN=1$ quiver diagram for the $\mathbb{C}^3/\mathbb{Z}_n$ theory. The quiver is a circle of $n$ nodes.}
\label{fig:quivc3zn}
\end{figure}

We now consider the worldvolume gauge theory of $N$ D3 branes probing a $\mathbb{C}^3/\mathbb{Z}_n$ singularity, with the $\mathbb{Z}_n$ action on the coordinates of $\mathbb{C}^3$ given by 
\begin{equation}
\mathbb{Z}_n = \left\{
\begin{pmatrix}
\omega_n^k & & \\
&\omega_n^{k} & \\
& & \omega_n^{-2k}
\end{pmatrix}, 1 \leq k \leq n
\right\}.
\label{eq:c3znact}
\end{equation}
as studied in \cite{Douglas:1997de}. 
This theory has the quiver diagram in figure \ref{fig:quivc3zn} and superpotential
\begin{align}
W &= \sum_{i=1}^n \lp A_{i,i+1}B_{i+1,i+2} - B_{i,i+1}A_{i+1,i+2} \rp C_{i+2,i}. 
\end{align}
The F-term relations are
\begin{align}
A_{i,i+1}B_{i+1,i+2} = B_{i,i+1}A_{i+1,i+2}, \nonumber \\
B_{i+1,i+2}C_{i+2,i} = C_{i+1,i-1}B_{i-1,i}, \nonumber \\
A_{i+1,i+2}C_{i+2,i} = C_{i+1,i-1}A_{i-1,i}.
\end{align}
%%%%%%%%%%%%%%%%%%%%%%%%%%%%%%%%%%%%%%%%%%%%%%%%%%%%%%%%%%%%%%%%%%%
\subsection{\texorpdfstring{$N=1$}{N=1} Moduli Space}
\label{subsec:c3znmodsp}
%%%%%%%%%%%%%%%%%%%%%%%%%%%%%%%%%%%%%%%%%%%%%%%%%%%%%%%%%%%%%%%%%%%
%%%%%%%%%%%%%%%%%%%%%%%%%%%%%%%%%%%%%%%%%%%%%%%%%%%%%%%%%%%%%%%%%%%
To find the moduli space of this theory, we work with a set of gauge-invariant monomials similar to those of the previous section. However, due to the structure of the quiver, we must use slightly different sets for even and odd $n$. Useful coordinates are given by
\begin{equation}
X_{a,b,c} = \prod_{i=1}^aA_{i,i+1} \prod_{j=1}^bB_{a+j,a+j+1} \prod_{k=1}^c C_{a+b+1+(k-1)(n-2),a+b+1+k(n-2)}
\end{equation}
so that essentially, $X_{a,b,c} \sim A^a B^b C^c$ and all subscripts are modulo $n$. These parameters are subject to the relation $X_{a_1,b_1,c_1}X_{a_2,b_2,c_2} = X_{a_1+a_2,b_1+b_2,c_1+c_2}$. 
The gauge-invariants for $n$ odd are
\begin{center}
\begin{tabular}{c c c c}
$X_{n,0,0} $ & $X_{2,0,1} $ & $X_{1,0,\frac{n+1}{2}}$&$X_{0,0,n}$\nonumber \\
$X_{n-1,1,0} $& $X_{1,1,1}$ & $X_{0,1,\frac{n+1}{2}}$ \nonumber \\
$\vdots$ & $X_{0,2,1}$ \nonumber \\
$X_{0,n,0} $ \nonumber 
\end{tabular}
\end{center} 
The space spanned by these coordinates subject to $X_{a_1,b_1,c_1}X_{a_2,b_2,c_2} = X_{a_1+a_2,b_1+b_2,c_1+c_2}$ is the space $\bC^3 / \bZ_n$ with odd $n$.
For even $n$, we use
\begin{center}
\begin{tabular}{c c c}
$X_{n,0,0} $ & $X_{2,0,1} $ & $X_{0,0,n/2}$\nonumber \\
$X_{n-1,1,0} $& $X_{1,1,1}$ & $Y_{0,0,n/2}$ \nonumber \\
$\vdots$ & $X_{0,2,1}$ \nonumber \\
$X_{0,n,0} $ \nonumber 
\end{tabular}
\end{center}
where $X_{a,b,c}$ is as above, and we have added $X_{0,0,n/2} = C_{1,n-1} \dots C_{3,1}$ and $Y_{0,0,n/2} = C_{2,n} \dots C_{4,2}$. There is now an additional relation $X_{a,b,c}Y_{0,0,n/2}^m \sim X_{a,b,c+nm/2}$ when $a+b>0$. 

The additional coordinate $Y_{0,0,n/2}$ is required to account for the fact that there are two distinct gauge-invariant operators with $n/2$ $C$'s. The moduli space for even $n$ has two different branches:
\begin{enumerate}
\item $\{X_{n,0,0}, \dots, X_{0,n,0}, X_{2,0,1}, X_{1,1,1}, X_{0,2,1}, X_{0,0,n/2}, Y_{0,0,n/2}\}$ subject to $X_{a_1,b_1,c_1}X_{a_2,b_2,c_2} $ $= X_{a_1+a_2,b_1+b_2,c_1+c_2}$ and $X_{0,0,n/2}=Y_{0,0,n/2}$,
\item $\{X_{n,0,0}, \dots, X_{0,n,0}, X_{2,0,1}, X_{1,1,1}, X_{0,2,1}, X_{0,0,n/2}, Y_{0,0,n/2}\}$ subject to $X_{a,b,c} = 0$ when $a+b>0$.
\end{enumerate}
We see that, as with the $\bC^3 / \bZ_2$ theory, there are multiple branches of moduli space. Specifically, there is one main branch of the moduli space which is $\bC^3 / \bZ_n$ and an extra branch which is $\bC^2$, with coordinates $X_{0,0,n/2}$ and $Y_{0,0,n/2}$. We denote this full moduli space $\bC^3 / \bZ_n \cup \bC^2$, where the union is defined so that the spaces share the line defined by $X_{0,0,n/2} = Y_{0,0,n/2}$ and $X_{a,b,c}=0$ for $a+b>0$.
%%%%%%%%%%%%%%%%%%%%%%%%%%%%%%%%%%%%%%%%%%%%%%%%%%%%%%%%%%%%%%%
\subsection{\texorpdfstring{$W=0$}{W=0} Large \texorpdfstring{$N$}{N} Chiral Ring}
\label{subsec:c3znw=0cr}
%%%%%%%%%%%%%%%%%%%%%%%%%%%%%%%%%%%%%%%%%%%%%%%%%%%%%%%%%%%%%%%
%%%%%%%%%%%%%%%%%%%%%%%%%%%%%%%%%%%%%%%%%%%%%%%%%%%%%%%%%%%%%%%
Using the P\'{o}lya enumeration theorem, the generating function for single-trace operators is of the form
\begin{equation}
F^{(\infty)}_S \lp x_i \rp = - \sum_{k=1}^{\infty}\frac{\varphi(k)}{k}\log\left[ 1-f \lp x_i^k \rp \right]
\end{equation}
and the multi-trace operator generating function is of the form
\begin{equation}
F^{(\infty)}_M \lp x_i \rp = \prod_{k=1}^{\infty} \frac{1}{1-f \lp x_i^k \rp }.
\end{equation}
for some colour generating function $f(x_i)$. This function has a term for every closed loop in the quiver. In combinatorics language, each closed loop constitutes a colour. We can make necklaces (single-trace operators) by combining beads of different colours (closed loops of operators) in a necklace (trace). However, in our colour generating function we must subtract the product of any two loops that do not overlap and thus can not be placed beside each other in a necklace. This means that we subtract the contributions from operators where there is a product over two loops that do not share a node. 

Similarly, it is necessary add terms to $f \lp x_i^k \rp$ that are cubic in non-intersecting loops, subtract terms that are quartic, and so on. As an example, consider the colour generating function for 3-ary necklaces of beads where none of the colours can be placed beside each other. This is simply the sum of three generating functions of 1-ary necklaces:
\begin{align}
F^{(\infty)}_S \lp x_,y,z \rp &= - \sum_{n=1}^{\infty}\frac{\varphi(d)}{d}\log\left[ 1- x^n \right]+\log\left[ 1- y^n \right]+\log\left[ 1- z^n \right] \nonumber \\
&= - \sum_{n=1}^{\infty}\frac{\varphi(d)}{d} \log\left[ 1- (x^n+y^n+z^n-x^ny^n - x^nz^n - y^nz^n + x^ny^nz^n) \right].
\end{align}
The colour generating function is then
\begin{equation}
f(x,y,z) = x+y+z-xy - xz - yz + xyz.
\end{equation}
As we go to higher number of colours, we must continue this pattern of addition and subtraction. This is in agreement with the formula found in \cite{Mattioli:2014yva}:
\begin{equation}
F^{(\infty)}_M (x,y,z) = \pe [F^{(\infty)}_S(x,y,z)] = \prod_i \frac{1}{\text{det}(\mathbb{I}-X_n(x^i,y^i,z^i))},
\end{equation}
where $X_n$ is the weighted adjacency matrix for the graph described by the quiver.
%%%%%%%%%%%%%%%%%%%%%%%%%%%%%%%%%%%%%%%%%%%%%%%%%%%%%%%%%%%%%%%
\subsection{\texorpdfstring{$W \neq 0$}{W neq 0} Large \texorpdfstring{$N$}{N} Chiral Ring}
\label{subsec:c3znwneq0cr}
%%%%%%%%%%%%%%%%%%%%%%%%%%%%%%%%%%%%%%%%%%%%%%%%%%%%%%%%%%%%%%%
%%%%%%%%%%%%%%%%%%%%%%%%%%%%%%%%%%%%%%%%%%%%%%%%%%%%%%%%%%%%%%%

We now describe the generating functions for even and odd $n$. As these derivations are rather lengthy, we relegate them to  appendix \ref{app:genc3zn} and here simply summarize the results. For general odd $n$ the generating function is
\begin{equation}
\boxed{F^{(\infty)}_S \lp a,b,c \rp = \left[ \sum_{m=0} ^{\infty} c^{nm} \right] \left[ \sum_{j=0}^{\frac{n-1}{2}}  \sum_{\ell = 0} ^{\infty} \sum_{k=0}^{n \ell + 2j} a^{k}b^{n \ell + 2j - k}c^j + c^{\frac{n+1}{2}} \sum_{j=0}^{\frac{n-3}{2}}  \sum_{\ell=0}^{\infty} \sum_{k=0}^{n \ell + 2j + 1} a^{k} b^{n \ell + 2j+1-k}c^j \right]}.
\label{eq:fsc3zn}
\end{equation}
which has the rational form
\begin{align}
F^{(\infty)}_S \lp a,b,c \rp =& \frac{1}{b-a} \frac{1}{1-c^n} \Bigg[ \frac{b \left(-b^n c^n-\left(b^2 c\right)^{\frac{n+1}{2}}+b c^{\frac{n+1}{2}}+1\right)}{\left(1-b^2 c\right) \left(1-b^n\right)} \nonumber \\
& \phantom{\frac{1}{b-a} \frac{1}{1-c^n} \bigg[}-\frac{a \left(-a^n c^n-\left(a^2 c\right)^{\frac{n+1}{2}}+a
   c^{\frac{n+1}{2}}+1\right)}{\left(1-a^2 c\right) \left(1-a^n\right)} \Bigg].
\label{eq:fsc3znrat}
\end{align}
We can then get the multi-trace operator generating function using plethystics:
\begin{equation}
\boxed{F^{(\infty)}_M(a,b,c) = \prod_{m=0}^{\infty} \left[ \prod_{j=0}^{\frac{n-1}{2}} \prod_{\ell=0}^{\infty} \prod_{k=0}^{n \ell + 2j} \frac{1}{1-c^{nm+j}a^k b^{n \ell + 2j - k}}\right]\left[ \prod_{j=0}^{\frac{n-3}{2}} \prod_{\ell=0}^{\infty} \prod_{k=0}^{n \ell + 2j + 1} \frac{1}{1-c^{nm+\frac{n+1}{2}+j}a^k b^{n \ell + 2j - k}} \right]}.
\end{equation}
Taking $a \rightarrow t$, $b \rightarrow t$, $c \rightarrow t$ in equation \eqref{eq:fsc3zn}, we get 
\begin{equation}
F^{(\infty)}_S \lp t \rp = \frac{-t^{2 n}-n t^{n+3}-t^{2 n+3}-2 t^{\frac{3 (n+1)}{2}}+2 t^{\frac{n+3}{2}}+n t^n+t^3+1}{\left(t^3-1\right)^2 \left(t^n-1\right)^2},
\label{eq:fsc3znt}
\end{equation}
which agrees with the result which can be calculated using the methods in \cite{Benvenuti:2006qr} for general odd $n$. 

For even $n$ the generating function is
\begin{equation}
\boxed{F^{(\infty)}_S (a,b,c) = \sum_{m=0} ^{\infty} c^{\frac{nm}{2}} \sum_{j=0}^{\frac{n}{2}-1} \sum_{\ell =0}^{\infty} \sum_{k=0}^{n \ell + 2j } a^{n \ell + 2j -k} b^k c^j + \sum_{m=1}^{\infty} c^{\frac{nm}{2}}},
\label{eq:fsc3zneven}
\end{equation}
which has the rational form
\begin{equation}
F^{(\infty)}_S(a,b,c) = \frac{1}{(a-b)
   \left(1-c^{n/2}\right)}\left[\frac{a \left(1-\left(a^2 c\right)^{n/2}\right)}{\left(1-a^2 c\right) \left(1-a^n\right)}-\frac{b \left(1-\left(b^2 c\right)^{n/2}\right)}{\left(1-b^2 c\right) \left(1-b^n\right)}\right]+\frac{c^{n/2}}{1-c^{n/2}}.
\end{equation}
The multi-trace operator generating function is then
\begin{equation}
\boxed{F^{(\infty)}_M (a,b,c) = \left[\prod_{m=0} ^{\infty} \prod_{j=0}^{\frac{n}{2}-1} \prod_{\ell =0}^{\infty} \prod_{k=0}^{n \ell + 2j } \frac{1}{1-a^{n \ell + 2j -k} b^k c^{j+\frac{nm}{2}}}\right] \left[\prod_{m=1}^{\infty} \frac{1}{1-c^{\frac{nm}{2}}}\right]}.
\label{eq:FMc3znodd}
\end{equation}
Taking $a \rightarrow t$, $b \rightarrow t$, $c \rightarrow t$ in equation \eqref{eq:fsc3zneven} yields
\begin{equation}
F^{(\infty)}_S(t) = \frac{\frac{n \left(1-\left(t^3\right)^{n/2}\right) t^n}{\left(1-t^3\right) \left(1-t^n\right)^2}+\frac{1-\left(t^3\right)^{n/2}}{\left(1-t^3\right) \left(1-t^n\right)}+\frac{2
   \left(-\frac{1}{2} n t^{3 n/2}+\left(\frac{n}{2}-1\right) t^{\frac{3 n}{2}+3}+t^3\right)}{\left(1-t^3\right)^2 \left(1-t^n\right)}}{1-t^{n/2}} + \frac{t^{n/2}}{1-t^{n/2}}.
\label{eq:fsc3znevent}
\end{equation}
While the first term is the one that can be found using the methods in \cite{Benvenuti:2006qr}, the second term is new.

%%%%%%%%%%%%%%%%%%%%%%%%%%%%%%%%%%%%%%%%%%%%%%%%%%%%%%%%%%%%%%%
\subsubsection{\texorpdfstring{$U(\infty)$}{U(infty)} Fock Space}
\label{subsec:c3znfock}
%%%%%%%%%%%%%%%%%%%%%%%%%%%%%%%%%%%%%%%%%%%%%%%%%%%%%%%%%%%%%%%
%%%%%%%%%%%%%%%%%%%%%%%%%%%%%%%%%%%%%%%%%%%%%%%%%%%%%%%%%%%%%%%
We can see from equation \eqref{eq:FMc3znodd} that the generating function for multi-trace operators in the large $N$ $\bC^3 / \bZ_n$ theory with even $n$ is equal to the generating function for the Fock space of bosons on the transverse space times an extra factor. More specifically, it is
\begin{equation}
F^{(\infty)}_M(a,b,c) = F_{\text{Fock}}(\bC^3 / \bZ_n) \times F_{\text{Fock}}(\bC).
\end{equation}
Similarly to the $\bC^3 / \bZ_2$ theory, this is the generating function for the Fock space of bosons moving on $\bC^3 / \bZ_n \coprod \bC$, where $\coprod$ indicates a disjoint union. For odd $n$ we do not have this extra factor, and the multi-trace operator generating function is equal to the generating function for the Fock space on $\bC^3 / \bZ_n$.
%%%%%%%%%%%%%%%%%%%%%%%%%%%%%%%%%%%%%%%%%%%%%%%%%%%%%%%%%%%%%%%
\subsection{\texorpdfstring{$N=1$}{N=1} Chiral Ring}
\label{subsec:c3znu1cr}
%%%%%%%%%%%%%%%%%%%%%%%%%%%%%%%%%%%%%%%%%%%%%%%%%%%%%%%%%%%%%%%
%%%%%%%%%%%%%%%%%%%%%%%%%%%%%%%%%%%%%%%%%%%%%%%%%%%%%%%%%%%%%%%
For the case of odd $n$ there is a one-to-one mapping between operators in the $U(1)$ theory and single-trace operators in the large $N$ theory; this mapping is $X_{a,b,c} \rightarrow \Tr(X_{a,b,c})$. Because of this the generating function for the $U(1)$ chiral ring for this theory is simply the generating function for single-trace operators in the large $N$ theory, given in equation \eqref{eq:fsc3zn}. Here the chiral ring has the basis
\begin{equation}
\{X_{a,b,c} | a+b+c(n-2) \equiv 0 \mod n \},
\end{equation}
with $X_{a,b,c}$ as in section \ref{subsec:c3znmodsp}.
This is the ring generated by the gauge-invariant operators
\begin{center}
\begin{tabular}{c c c c}
$X_{n,0,0} $ & $X_{2,0,1} $ & $X_{1,0,\frac{n+1}{2}}$&$X_{0,0,n}$\nonumber \\
$X_{n-1,1,0} $& $X_{1,1,1}$ & $X_{0,1,\frac{n+1}{2}}$ \nonumber \\
$\vdots$ & $X_{0,2,1}$ \nonumber \\
$X_{0,n,0} $ \nonumber 
\end{tabular}
\end{center}

For even $n$, things are slightly different. The set of elements in the $U(1)$ chiral ring for this theory is not equal to the set of single-trace operators in the large $N$ theory; this is because for $N>1$ the $X_{0,0,n/2}$ and $Y_{0,0,n/2}$ operators transform in the adjoint representations of the $U(N)_1$ and $U(N)_2$ gauge groups, respectively. Thus we cannot have single-trace operators of the form $\Tr(X_{0,0,n/2}^{m_1}Y_{0,0,n/2}^{m_2})$ in the large $N$ theory, though in the $U(1)$ theory operators of the form $X_{0,0,n/2}^{m_1}Y_{0,0,n/2}^{m_2}$ are allowed. This means that there is {\it not} a one-to-one mapping between operators in the $U(1)$ theory and single-trace operators in the large $N$ theory.

For even $n$, the $U(1)$ chiral ring has the basis
\begin{equation}
\{X_{a,b,c} | a+b+c(n-2) \equiv 0 \mod n \} \cup \{ X_{0,0,nm_1/2} Y_{0,0,nm_2/2}| m_1 \in \bZ_{\geq 0} ,m_2 \in \mathbb{Z}_+ \},
\end{equation}
modulo the equivalence $X_{a_1,b_1,c_1}X_{a_2,b_2,c_2} \sim X_{a_1+a_2,b_1+b_2,c_1+c_2}$ and $X_{a,b,c}Y_{0,0,nm/2} \sim X_{a,b,c+nm/2}$ when $a+b>0$, with $Y_{0,0,nm/2}=Y_{0,0,n/2}^m$ and $X_{a,b,c}$ and $Y_{0,0,n/2}$ as defined in section \ref{subsec:c3znmodsp}. Again we see that the chiral ring is equal to the ring of holomorphic polynomials on the moduli space, $\bC^3 / \bZ_n \cup \bC^2$ and the counting function is given by
\begin{equation}
F^{(1)} (a,b,c) = \sum_{m=0} ^{\infty} \sum_{j=0}^{\frac{n}{2}-1} \sum_{\ell =0}^{\infty} \sum_{k=0}^{n \ell + 2j } a^{n \ell + 2j -k} b^k c^{j+\frac{nm}{2}} + \sum_{m_1=0}^{\infty} \sum_{m_2=0}^{\infty} c^{\frac{n}{2}(m_1+m_2)} - \sum_{m=0}^{\infty} c^{\frac{n}{2}m}.
\label{eq:c3znFSu1}
\end{equation}

As was the case in the $\bC^3 / \bZ_2$ example, this generating function is a sum of three terms. The first is the generating function for the ring of holomorphic polynomials on $\bC^3 / \bZ_n$. The second is the generating function for the ring of holomorphic polynomials on $\bC^2$. Finally, the last term subtracts the generating function for the ring of holomorphic polynomials on the intersection. This generating function then is the generating function for holomorphic polynomials on the moduli space $\bC^3 / \bZ_n \cup \bC^2$.

As was the case with the $\bC^3 / \bZ_2$ theory we can go from the chiral ring, $\cR = \cR_0 / \cI_0$, to the ring of holomorphic polynomials on the two branches of moduli space by quotienting by two ideals. In the case of the $\bC^3 / \bZ_n$ theory with even $n$ the two ideals are $\cI_1 = \langle (X_{0,0,n/2} - Y_{0,0,n/2}) \rangle$ and $\cI_2 = \langle X_{n,0,0}, X_{n-1,1,0}, \dots, X_{0,n,0}, X_{2,0,1}, X_{1,1,1}, X_{0,2,1} \rangle$. Alternatively, we could have obtained these two rings from the $W=0$ chiral ring using minimal prime ideals. The ideal $\cI_0$ is not a prime ideal and there are two minimal prime ideals, $\cI_1'$ and $\cI_2'$ over $\cI_0$. Quotienting $\cR_0$ by these two ideals gives the ring of functions on $\bC^3 / \bZ_n$ and $\bC^2$. This is analogous to the situation depicted in figure \ref{fig:ideals}.
%%%%%%%%%%%%%%%%%%%%%%%%%%%%%%%%%%%%%%%%%%%%%%%%%%%%%%%%%%%%%%%
\subsection{Conclusion}
\label{subsec:c3zncon}
%%%%%%%%%%%%%%%%%%%%%%%%%%%%%%%%%%%%%%%%%%%%%%%%%%%%%%%%%%%%%%%
%%%%%%%%%%%%%%%%%%%%%%%%%%%%%%%%%%%%%%%%%%%%%%%%%%%%%%%%%%%%%%%
In this section we have attempted to find out how the $\bC^3 / \bZ_n$ theory fits into our description of moduli spaces, chiral rings, and Fock spaces of branes. For $n$ odd, the $\bC^3 / \bZ_n$ theory fits into this story in the same way as $\cN=4$  and the conifold. The moduli space is simply $\bC^3 / \bZ_n$ and the set of elements in the $U(1)$ chiral ring is the set of functions on $\bC^3 / \bZ_n$. This set is equal to the set of single-trace operators, and the multi-trace operator generating function is equal to the generating function for the Fock space of bosons moving on $\bC^3 / \bZ_n$.

For $n$ even, the story needs to be amended slightly:
\begin{enumerate}
\item The moduli space has one main branch where the space is simply $\bC^3 / \bZ_n$, however it also has an extra branch where the space is $\bC^2$.
\item The set of elements in the $U(1)$ chiral ring is then not quite the set of functions on $\bC^3 / \bZ_n$, but rather the set of functions on $\bC^3 / \bZ_n \cup \bC^2$.
\item The set of elements in the $U(1)$ chiral ring is also not equal to the set of single-trace operators in the large $N$ theory.
\item The multi-trace operator generating function is equal to the generating function for the Fock space for bosons on $\bC^3 / \bZ_n \coprod \bC$.
\end{enumerate}
%%%%%%%%%%%%%%%%%%%%%%%%%%%%%%%%%%%%%%%%%%%%%%%%%%%%%%%%%%%%%%%%%%%
\section{\texorpdfstring{$\bC^3 / \hA_N$}{C3/An}}
\label{sec:c3an}
%%%%%%%%%%%%%%%%%%%%%%%%%%%%%%%%%%%%%%%%%%%%%%%%%%%%%%%%%%%%%%%%%%%
%%%%%%%%%%%%%%%%%%%%%%%%%%%%%%%%%%%%%%%%%%%%%%%%%%%%%%%%%%%%%%%%%%%
We now consider the $\bC^3 / \hA_n$ theory studied in \cite{Douglas:1996sw,Kachru:1998ys}, with the action of $\hA_n$ defined by \begin{equation}
\mathbb{Z}_n = \left\{
\begin{pmatrix}
\omega_n^k & & \\
&\omega_n^{-k} & \\
& & 1
\end{pmatrix}, 1 \leq k \leq n
\right\}.
\end{equation}
\begin{figure}[ht]
\centering
\includegraphics[scale=0.2]{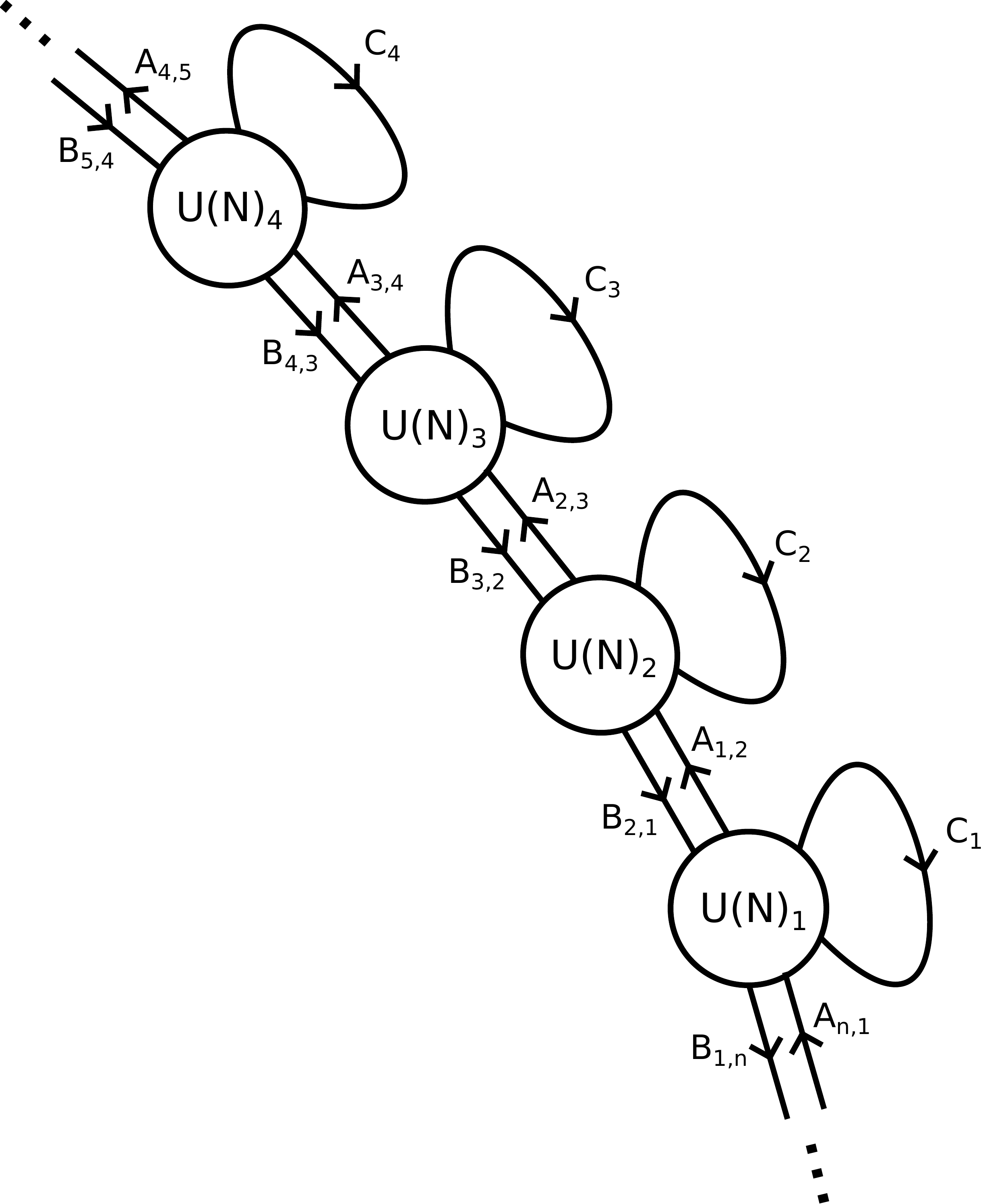}
\caption{The $\cN=1$ quiver diagram for the the $\mathbb{C}^3/ \hat{A}_n$ theory.}
\label{fig:quivc2an}
\end{figure}
The $\cN=1$ quiver diagram for this theory is as shown in figure \ref{fig:quivc2an},  and the superpotential is
\begin{equation}
W = \sum_{i=1}^n C_i\left( A_{i,i+1}B_{i+1,i} - B_{i,i-1}A_{i,i-1} \right).
\end{equation}
The F-term relations are
\begin{align}
A_{i,i+1}B_{i+1,i} &= B_{i,i-1}A_{i-1,i}, \nonumber \\
B_{i+1,i}C_{i} &= C_{i+1}B_{i+1,i}, \nonumber \\
A_{i,i+1}C_{i+1} &= C_{i}A_{i,i+1}.
\end{align}
The problem of counting multi-trace operators for this theory along with other $\cN=2$ theories was also considered in \cite{Hanany:2006uc} where they derive the Higgs and Coulomb branch generating functions separately and then combine them.
%%%%%%%%%%%%%%%%%%%%%%%%%%%%%%%%%%%%%%%%%%%%%%%%%%%%%%%%%%%%%%%%%%%
\subsection{\texorpdfstring{$N=1$}{N=1} Moduli Space}
\label{subsec:c3anmodsp}
%%%%%%%%%%%%%%%%%%%%%%%%%%%%%%%%%%%%%%%%%%%%%%%%%%%%%%%%%%%%%%%%%%%
%%%%%%%%%%%%%%%%%%%%%%%%%%%%%%%%%%%%%%%%%%%%%%%%%%%%%%%%%%%%%%%%%%%
The moduli space of this theory is parametrised by the operators $X=A_{1,2}A_{2,3} \dots A_{n,1}$, $Y=B_{1,n}B_{n,n-1} \dots B_{2,1}$, $Z=A_{1,2}B_{2,1}$, $C_1$,..., $C_{n-1}$, and $C_n$, subject to the relation $XY=Z^n$.
As was the case for the $\bC^3 / \bZ_n$ theory with even $n$, there are two branches of solutions:
\begin{enumerate}
\item $\{X,Y,Z, C_1, \dots , C_n\}$ subject to $XY=Z^n$, $C_1=C_2= \dots =C_n$.
\item $\{X,Y,Z, C_1, \dots , C_n\}$ subject to $X=Y=Z=0$.
\end{enumerate}
The first branch is simply $\bC^3 / \hA_n$, and the second branch is $\bC^n$. Again we see the presence of one main branch identical to the transverse space to the D3-branes as well as an extra branch. We denote this space as $\bC^3 / \hA_n \cup \bC^n$, where the union is such that the two branches share the line $X = Y = Z = 0$, $C_1 = C_2 = \dots = C_n$. The picture for this moduli space is as in figure \ref{fig:c3z2mod}, except that now the extra branch is $n$-dimensional.

The first branch of this moduli space is a mixed branch since it contains both Higgs and Coulomb branch operators. The second branch however is a purely Coulomb branch. The first branch contains as a sub-space the $C_i=0$, pure Higgs branch.
%%%%%%%%%%%%%%%%%%%%%%%%%%%%%%%%%%%%%%%%%%%%%%%%%%%%%%%%%%%%%%%
\subsection{\texorpdfstring{$W=0$}{W=0} Large \texorpdfstring{$N$}{N} Chiral Ring}
\label{subsec:c3anw=0cr}
%%%%%%%%%%%%%%%%%%%%%%%%%%%%%%%%%%%%%%%%%%%%%%%%%%%%%%%%%%%%%%%
%%%%%%%%%%%%%%%%%%%%%%%%%%%%%%%%%%%%%%%%%%%%%%%%%%%%%%%%%%%%%%%
As in the previous section the single- and multi-trace operator generating functions are
\begin{align}
F^{(\infty)}_S \lp x_i \rp &= - \sum_{n=1}^{\infty}\frac{\varphi(d)}{d}\log\left[ 1-f \lp x_i^n \rp \right], \nonumber
\\
F^{(\infty)}_M \lp x_i \rp &= \prod_{n=1}^{\infty} \frac{1}{1-f \lp x_i^n \rp },
\end{align}
where the colour generating function is determined in the previous section, matching the results of \cite{Mattioli:2014yva}. 

%%%%%%%%%%%%%%%%%%%%%%%%%%%%%%%%%%%%%%%%%%%%%%%%%%%%%%%%%%%%%%%
\subsection{\texorpdfstring{$W \neq 0$}{W neq 0} Large \texorpdfstring{$N$}{N} Chiral Ring}
\label{subsec:c3anwneq0cr}
%%%%%%%%%%%%%%%%%%%%%%%%%%%%%%%%%%%%%%%%%%%%%%%%%%%%%%%%%%%%%%%
%%%%%%%%%%%%%%%%%%%%%%%%%%%%%%%%%%%%%%%%%%%%%%%%%%%%%%%%%%%%%%%
When we have a non-zero superpotential the F-term relations tell us that any single-trace operator with $n_1$ $A$'s, $n_2$ $B$'s and $n_3$ $C$'s is equivalent to any other single-trace operator with $n_1$ $A$'s, $n_2$ $B$'s and $n_3$ $C$'s. The only exception to this rule is when $n_1=n_2=0$. In that case there are $n$ different operators we can have of the form $\Tr (C_i^{n_3})$. For the first set of operators we can construct any gauge-invariant operator by inserting $A_{1,2}B_{2,1}$, $A_{1,2} \dots A_{n,1}$, $B_{1,n}, \dots B_{2,1}$ and $C_1$ into a trace recursively. This means that the generating function will be
\begin{equation}
\boxed{F^{(\infty)}_S(a,b,c) = \sum_{m=0}^{n-1} \sum_{j=0}^{\infty} a^{nj+m} \sum_{k=0}^{\infty}c^k \sum_{l=0}^{\infty} b^{nl + m} + (n-1) \sum_{k=1}^{\infty} c^k},
\label{eq:c3anFS}
\end{equation}
with rational form
\begin{equation}
F^{(\infty)}_S(a,b,c) = \frac{1-(ab)^n}{(1-a^n)(1-b^n)(1-c)(1-ab)} + \frac{(n-1)c}{1-c}.
\end{equation}
Taking $a,b,c \rightarrow t$ gives
\begin{equation}
F^{(\infty)}_S(t) = \frac{1+t^n}{(1-t^n)(1-t)(1-t^2)} + \frac{(n-1)t}{1-t^n}.
\end{equation}
The first term of this formula matches the result given in \cite{Benvenuti:2006qr}. The presence of the second term is caused by considering the operators $\Tr(C_i^k)$ and $\Tr(C_j^k)$ to be inequivalent for $i \neq j$ and is considered in \cite{Hanany:2006uc}.

We can use plethystics once again to get the multi-trace operator generating function:
\begin{equation}
\boxed{F^{(\infty)}_M(a,b,c) = \left[\prod_{m=0}^{n-1} \prod_{j=0}^{\infty}  \prod_{k=0}^{\infty} \prod_{l=0}^{\infty} \frac{1}{1-a^{nj+m}b^{nl + m}c^k}\right] \left[ \sum_{k=1}^{\infty} \frac{1}{1-c^k}\right]^{n-1}}.
\label{eq:c3anFM}
\end{equation}
%%%%%%%%%%%%%%%%%%%%%%%%%%%%%%%%%%%%%%%%%%%%%%%%%%%%%%%%%%%%%%%
\subsubsection{\texorpdfstring{$U(\infty)$}{U(infty)} Fock Space}
\label{subsubsec:c3anfock}
%%%%%%%%%%%%%%%%%%%%%%%%%%%%%%%%%%%%%%%%%%%%%%%%%%%%%%%%%%%%%%%
%%%%%%%%%%%%%%%%%%%%%%%%%%%%%%%%%%%%%%%%%%%%%%%%%%%%%%%%%%%%%%%
From equation \eqref{eq:c3anFS} one can see that the large $N$ set of single-trace operators is not equal to set of elements in the ring of holomorphic polynomials on $\bC^3 / \hA_n$. As a consequence of this, the multi-trace operator generating function is not equal to the generating function for the Fock space of bosons on $\bC^3 / \hA_n$. Instead, as can be seen from equation \eqref{eq:c3anFM} we have
\begin{equation}
F^{(\infty)}_M(a,b,c) = F_{\text{Fock}}(\bC^3 / \hA_n) \times \left[F_{\text{Fock}}(\bC)\right]^{n-1}.
\end{equation}
This is the generating function for the multi-particle Fock space of bosons moving on $\bC^3 / \hA_n \left[ \coprod \bC \right]^{n-1}$, {\it i.e.},~it is the Fock space for bosons that can have wavefunctions either in the ring of holomorphic polynomials on $\bC^3 / \hA_n$ or in the ring of holomorphic polynomials on any of $n-1$ copies of $\bC$.
%%%%%%%%%%%%%%%%%%%%%%%%%%%%%%%%%%%%%%%%%%%%%%%%%%%%%%%%%%%%%%%
\subsection{\texorpdfstring{$N=1$}{N=1} Chiral Ring}
\label{subsec:c3anu1cr}
%%%%%%%%%%%%%%%%%%%%%%%%%%%%%%%%%%%%%%%%%%%%%%%%%%%%%%%%%%%%%%%
%%%%%%%%%%%%%%%%%%%%%%%%%%%%%%%%%%%%%%%%%%%%%%%%%%%%%%%%%%%%%%%
As was the case in the previous two sections the set of elements in the $N=1$ chiral ring for the $\bC^3 / \hA_n$ theory differs from the set of single trace operators in the large $N$ theory. This is because there are operators of the form $C_1^{k_1} \dots C_n^{k_n}$ in the $N=1$ theory whose analogue in the set of large $N$ single-trace operators, $\Tr (C_1^{k_1} \dots C_n^{k_n})$ do not exist.

So, the chiral ring for the $\bC^3 / \hA_n$ theory has the following basis:
\begin{equation}
\{ X^{n_1} Y^{n_2} Z^{n_3} C_1^{n_4}| n_{1,2,4} \in \mathbb{Z}_{\geq 0}, n_3 \in [0,n-1] \cap \mathbb{Z} \} \cup \{ C_1^{m_1} C_2^{m_2} \cdots C_n^{m_n} | m_i \in \mathbb{Z}_{\geq 0}, \sum_{i=2}^{n}m_i>0 \},
\end{equation}
modulo the relation $X^{n_1} Y^{n_2} Z^{n_3} C_1^{m_4} C_2^{m_2} \cdots C_n^{m_n} \sim X^{n_1} Y^{n_2} Z^{n_3} C_1^{\sum_i m_i}$ when $n_1+n_2+n_3>0$. The counting function for this ring is
\begin{equation}
F^{(1)}(x,y,z,c) = \sum_{n_1=0}^{\infty} \sum_{n_2=0}^{\infty}  \sum_{n_4=0}^{\infty} \sum_{n_3=0}^{n-1} x^{n_1} y^{n_2} z^{n_3} c^{n_4} + \sum_{m_1=0}^{\infty} \dots \sum_{m_n=0}^{\infty} c^{m_1 + \dots m_n} - \sum_{m=1}^{\infty}c^{m}.
\label{eq:c3anFSu1}
\end{equation}
From this we can see once again that the generating function for the $N=1$ theory is the sum of the generating functions for the two branches minus the generating function for the intersection.This also matches the result found in \cite{Hanany:2006uc}.

We can go from the $N=1$ chiral ring, $\cR$, to the ring of functions on either of the two branches by quotienting by the ideals
\begin{equation}
\mathcal{I}_1 = \langle (C_2-C_1), (C_3-C_1), \cdots, (C_n-C_1) \rangle = \{ C_1^{m_1} \cdots C_n^{m_n} - C_1^{\sum_{i=1}^n m_i} | m_i \in \bZ_{\geq 0} \}.
\end{equation}
for $\bC^3 / \hA_n$ and
\begin{equation}
\mathcal{I}_2 = \langle X,Y,Z \rangle = \{ X^{n_1} Y^{n_2} Z^{n_3} C_1^{n_4}| n_{1,2,4} \in \mathbb{Z}_{\geq 0}, n_3 \in [0,n-1] \cap \mathbb{Z}, n_1+n_2+n_3>0 \}.
\end{equation}
for $\bC^n$. As with the other cases we also could have obtained these two rings by quotienting the $W=0$ chiral ring by the the two minimal prime ideals over $\cI_0$.

%%%%%%%%%%%%%%%%%%%%%%%%%%%%%%%%%%%%%%%%%%%%%%%%%%%%%%%%%%%%%%%
\subsection{Conclusion}
\label{subsec:c3ancon}
%%%%%%%%%%%%%%%%%%%%%%%%%%%%%%%%%%%%%%%%%%%%%%%%%%%%%%%%%%%%%%%
%%%%%%%%%%%%%%%%%%%%%%%%%%%%%%%%%%%%%%%%%%%%%%%%%%%%%%%%%%%%%%%
We have seen in this section another family of orbifold theories that have a very interesting relationship between transverse space, moduli space, $U(1)$ chiral ring and Fock space of branes. In particular we saw that
\begin{enumerate}
\item The moduli space of this theory has two separate branches: one main branch where the space is the same as the space transverse to the D3-branes, and one extra branch which is $\bC^n$.
\item The chiral ring of the $U(1)$ theory is equal to the ring of functions on the moduli space, $\bC^3 / \hA_n \cup \bC^n$, rather than the ring of functions on the transverse space, $\bC^3 / \hA_n $.
\item The $U(1)$ chiral ring has a set of elements that is not equal to the set of single-trace operators in the large $N$ theory.
\item The multi-trace operator generating function gave us the generating function for the Fock space of bosons moving on $\bC^3 / \hA_n \left[ \coprod \bC \right]^{n-1}$.
\end{enumerate}
%%%%%%%%%%%%%%%%%%%%%%%%%%%%%%%%%%%%%%%%%%%%%%%%%%%%%%%%%%%%%%%%%%%
\section{Conclusions}
\label{sec:outro}
%%%%%%%%%%%%%%%%%%%%%%%%%%%%%%%%%%%%%%%%%%%%%%%%%%%%%%%%%%%%%%%%%%%
%%%%%%%%%%%%%%%%%%%%%%%%%%%%%%%%%%%%%%%%%%%%%%%%%%%%%%%%%%%%%%%%%%%
In this work, we have described how certain seemingly simple relationships between chiral rings, moduli spaces, generating functions, and transverse geometries get modified in all but the most symmetric examples. While there is generally a main branch of moduli space which is the geometry transverse to the branes, or its symmetric products,  there are often  extra branches of moduli space.
This is a source of subtleties in the generating functions counting chiral ring operators. 
They also modify the na\"ive relationship between the generating function of multi-trace operators and the Fock space of bosons moving on the space transverse to the branes. 

There are a number of natural questions for future work. Since we have here only considered the $N=1$ and $N \rightarrow \infty$ limits, it would potentially be interesting to understand how our results are modified for finite $N>1$. The explicit derivation
of $ Sym^N ( X )$ from the chiral ring, where $X$ is the transverse space, is only known in a few cases. 
From our studies at $ N=1 $ and $ N \rightarrow \infty $, we expect this structure to be present, but with 
subtle modifications due to the extra branches. A systematic description and  derivation of this structure 
would be fascinating. Restricting attention to the $N=1$ and $ N \rightarrow \infty $ cases, is there a simple general 
mathematical/geometrical  formulation (bypassing explicit gauge theory computations), perhaps based on physical ideas around fractional branes, which can start from 
the chiral ring and moduli space at $N=1$ and derive  the Fock space structure at large $N$?
As we have seen in the examples, the Fock space structure we find from the gauge theory 
chiral operators always contains a factor which is the Fock space for the main branch corresponding 
to the geometry transverse to the 3-branes. However, while the extra Fock space factors are correlated 
with  the existence of extra branches, there is no simple rule like the existence of a Fock space at $ N \rightarrow \infty $ 
for every branch at $ N=1$. Is there a clear rule which replaces this naive rule? Even if such a rule existed, what would be the geometrical/mathematical meaning behind it?

One of the  motivations behind the present work was to ask whether there is a simple general 
algorithm  to deform the large $N$ counting formulae at zero superpotential to arrive at those for
non-zero superpotential. The latter have been the main focus of this work. 
The former admit simple general expressions based on the weighted adjacency matrix of 
 the quiver graph \cite{Pasukonis:2013ts,Mattioli:2014yva}. These expressions are also, somewhat surprisingly from 
 a physical point of view, related to some word-counting problems 
 based on the quiver \cite{Mattioli:2014yva}. We might hope that a deeper geometrical understanding of 
 the role of multiple branches at $ N=1$ and $ N \rightarrow \infty $ might provide 
 useful hints in finding such an algorithm.
 
 For the case of  D3-branes at the tip of a general toric Calabi-Yau cone,   all such theories can be reached via Higgsing $\bC^3/ \bZ_m \times \bZ_n$ theories, so it would nice to understand how the generating functions behave under the resulting flows. Similarly, one could hope to understand the relationship of generating functions in theories connected by (relevant) superpotential deformations. Although such deformations are in general complicated, and can lead to vastly different solutions to the F-terms, it might be possible to understand the effect of adding mass terms or other such very simple deformations. 

\pagebreak
%%%%%%%%%%%%%%%%%%%%%%%%%%%%%%%%%%%%%%%%%%%%%%%%%%%%%%%%%%%%%%%%%%%%%%
\begin{center}
\bf{Acknowledgements}
\end{center}
\medskip

We would like to thank David Berenstein, David Garner, Joseph Hayling, Yang-Hui He, Edward Hughes and Paolo Mattioli for useful discussions. This work is supported in part by the STFC Standard Grant ST/J000469/1 ``String Theory, Gauge Theory and Duality." JMG is supported by a Queen Mary University of London studentship.

\vspace*{0.2in}

%%%%%%%%%%%%%%%%%%%%%%%%%%%%%%%%%%%%%%%%%%%%%%%%%%%%%%%%%%%%%%%%%%%%%%
\appendix

%%%%%%%%%%%%%%%%%%%%%%%%%%%%%%%%%%%%%%%%%%%%%%%%%%%%%%%%%%%%%%%%%%%%%%
\section{Simplification of ST Generating Function for \texorpdfstring{$\cN=4$}{N=4} SYM With \texorpdfstring{$W \neq 0$}{W neq 0}}
\label{app:simp}
\renewcommand{\theequation}{A.\arabic{equation}}
\setcounter{equation}{0}
%%%%%%%%%%%%%%%%%%%%%%%%%%%%%%%%%%%%%%%%%%%%%%%%%%%%%%%%%%%%%%%%%%%%%%
\label{app:simpgenfun}
In section \ref{subsec:n=4wneq0cr} it was said that the generating function for $\cN=4$ SYM with non-zero superpotential could be derived using the P\'{o}lya enumeration theorem with $G=S_n$. We derive this now.

The symmetric group has
\begin{equation}
\frac{n!}{\prod_{k=1}^{n}(k!)^{j_k}}\prod_{k=1}^n\left(\frac{k!}{k}\right)^{j_k}\prod_{k=1}^n\frac{1}{j_k!}
\end{equation}
elements each with $j_k$ cycles of length $k$ for each partition $\{j_k\}$ of $n$. Thus the cycle index for the symmetric group is
\begin{equation}
Z_{S_n}\left(t_1,...,t_n\right) = \sum_{\{ j_k \} \vdash n } \frac{1}{\prod_{k=1}^{n}k^{j_k} \left(j_k!\right)} \prod_{k=1}^nt_k^{j_k}.
\end{equation}
The notation $\{j_k\} \vdash n$ means $\{j_k\}$ is a partition of $n$ so the sum is over partitions of $n$. This means that the generating function for large $N$ $\cN=4$ SYM single-trace operators with nonzero superpotential is
\begin{equation}
F^{(\infty)}_S \left(x,y,z\right) = \sum_{n=1}^{\infty} \sum_{\{ j_k \} \vdash n }\prod_{k=1}^n \frac{(x^k+y^k+z^k)^{j_k}}{k^{j_k} \left(j_k!\right)}.
\end{equation}
To show that this is equal to $(1-x)^{-1}(1-y)^{-1}(1-z)^{-1}$ we start by changing the sum over $n$ and the sum over partitions of $n$ to a infinite number of sums
\begin{align}
F^{(\infty)}_S \left(x,y,z\right) &= \sum_{n=1}^{\infty} \sum_{\{ j_k \} \vdash n }\prod_{k=1}^n \frac{1}{j_k!} \left(\frac{x^k+y^k+z^k}{k}\right)^{j_k} \nonumber \\
& = \left( \sum_{j_1=0}^{\infty}\sum_{j_2=0}^{\infty}... \right) \prod_{k=1}^n \frac{1}{j_k!} \left(\frac{x^k+y^k+z^k}{k}\right)^{j_k} \nonumber \\
& = \left( \sum_{j_1=0}^{\infty} \frac{1}{j_1!} \left(\frac{x+y+z}{1}\right)^{j_1} \right) \left( \sum_{j_2=0}^{\infty} \frac{1}{j_2!} \left(\frac{x^2+y^2+z^2}{2}\right)^{j_2} \right)... \nonumber \\
& = \prod_{k=1}^{\infty} \left[ \sum_{n=0}^{\infty} \frac{1}{n!}\left( \frac{x^k+y^k+z^k}{k} \right)^n \right] = \prod_{k=1}^{\infty}\exp\left( \frac{x^k+y^k+z^k}{k} \right) \nonumber \\
&= \exp \left( \sum_{k=1}^{\infty} \frac{x^k}{k} \right) \exp \left( \sum_{k=1}^{\infty} \frac{y^k}{k} \right) \exp \left( \sum_{k=1}^{\infty} \frac{z^k}{k} \right) = \frac{1}{1-x} \frac{1}{1-y} \frac{1}{1-z}
\end{align}
%%%%%%%%%%%%%%%%%%%%%%%%%%%%%%%%%%%%%%%%%%%%%%%%%%%%%%%%%%%%%%%%%%%%%%
\section{Derivation of \texorpdfstring{$\mathbb{C}^3/\mathbb{Z}_2$}{C3Z2} Molien Series}
\label{app:hanany}
\renewcommand{\theequation}{B.\arabic{equation}}
\setcounter{equation}{0}
%%%%%%%%%%%%%%%%%%%%%%%%%%%%%%%%%%%%%%%%%%%%%%%%%%%%%%%%%%%%%%%%%%%%%%
Here we calculate the single-trace operator generating function in the $\mathbb{C}^3/\mathbb{Z}_2$ theory using the methods given in \cite{Benvenuti:2006qr}. In \cite{Benvenuti:2006qr} it is said that the counting function for a $\mathbb{C}^3/G$ theory is given by the counting function for $\bC^3$ polynomials that are invariant under the action of the group $G$. It is also said that this is a classical problem and that the counting function is given by the Molien series:
\begin{equation}
M\lp t; G \rp = \frac{1}{|G|} \sum_{g \in G} \frac{1}{\det \lp \mathbb{I} -tg \rp}.
\end{equation}
Since $\mathbb{Z}_2$ has two elements whose action on the coordinates $x,y,z$ is given by the matrices
\begin{align}
\begin{pmatrix}
1&0&0 \\
0&1&0 \\
0&0&1
\end{pmatrix},
&&
\begin{pmatrix}
-1&0&0 \\
0&-1&0 \\
0&0&1
\end{pmatrix},
\end{align}
the Molien series is just
\begin{equation}
M\lp t; \mathbb{Z}_2 \rp = \frac{1}{2} \lp \frac{1}{(1-t)^3} + \frac{1}{(1-t)(1+t)^2} \rp = \frac{ \left(1+t^2\right)}{(1-t)^3 (1+t)^2}.
\end{equation}
This does not match the formula we have given in equation \eqref{eq:c3z2FSt}.
%%%%%%%%%%%%%%%%%%%%%%%%%%%%%%%%%%%%%%%%%%%%%%%%%%%%%%%%%%%%%%%%%%%%%%
\section{Derivation of Generating Functions for \texorpdfstring{$\mathbb{C}^3 / \mathbb{Z}_n$}{C3/Zn}}
\label{app:genc3zn}
\renewcommand{\theequation}{C.\arabic{equation}}
\setcounter{equation}{0}
%%%%%%%%%%%%%%%%%%%%%%%%%%%%%%%%%%%%%%%%%%%%%%%%%%%%%%%%%%%%%%%%%%%%%%

First let us consider single-trace operators that have no $C$ operators in them. These are all of the form $\Tr \lp A_{1,2}B_{2,3}A_{3,4}A_{4,5}...A_{ n-2, n-1}B_{ n-1, n}B_{n,1} \rp$. As required by gauge-invariance  the number of $A$'s and $B$'s in the trace will be $\ell n$, where $\ell$ is the number of loops we have traced around the quiver by following the arrows of the bifundamentals. The F-term equations allow us to interchange $A$'s and $B$'s freely (e.g. $A_{1,2}B_{2,3}=B_{1,2}A_{2,3}$) so that operators with the same number of $A$'s and the same number of $B$'s are equivalent.

When constructing a gauge-invariant single-trace operator from $A$'s and $B$'s the $A$ and $B$ operators move us from one node to the next and so we need $n$ of them to get back to the node we started at. The $C$ operators move us forward $n-2$ nodes. So for general $n$ the lowest lying single-trace operators have $\#A$'s $+\#B$'s $+\#C$'s $=3$ and are $\Tr \lp A_{1,2}A_{2,3}C_{3,1}\rp$, $\Tr \lp A_{1,2}B_{2,3}C_{3,1}\rp$, and $\Tr \lp B_{1,2}B_{2,3}C_{3,1}\rp$. The F-term relations tell us that any operator with $n_1$ $A$'s, $n_2$ $B$'s, and $n_3$ $C$'s is equivalent to any other operator with $n_1$ $A$'s, $n_2$ $B$'s, and $n_3$ $C$'s.

We can split the operators up into how many loops they form around the quiver. Consider the example of the $\mathbb{C}^3/\mathbb{Z}_5$ theory and, for the moment, neglect the existence of the $B$ operators. At one loop there are 2 operators:
\begin{align}
A^5, A^2C,
\end{align}
At two loops we have 4 operators:
\begin{align}
A^{10}, A^7C, A^4C^2, AC^3,
\end{align}
At three loops we have 6 operators:
\begin{align}
A^{15}, A^{12}C, A^9C^2, A^6C^3, A^3C^4, C^5.
\end{align}
And so on for higher number of loops.

Each time we have started with $\ell n$ $A$'s and gone from left to right by replacing $n-2=3$ $A$'s with a $C$. This means that the generating function will be
\begin{eqnarray}
F^{(\infty)}_S \lp a, c \rp &=& \lp 1 + a^5 + a^{10} + a^{15} + \dots \rp + c \lp a^2 + a^7 + a^{12} + \dots \rp + c^2 \lp a^4 + a^9 + \dots \rp \nonumber \\
&& + c^3 \lp a + a^6 + \dots \rp + c^4 \lp a^3 + a^8 + \dots \rp + c^5 \lp 1 + a^5 + a^{10} + \dots \rp + \dots \nonumber \\
&=& \left[ \sum_{m=0} ^{\infty} c^{nm} \right]\bigg[\lp 1 + a^5 + a^{10} + a^{15} + \dots \rp + c \lp a^2 + a^7 + a^{12} + \dots \rp \nonumber \\
&& \phantom{\left[ \sum_{m=0} ^{\infty} c^{nm} \right]\bigg[} + c^2 \lp a^4 + a^9 + \dots \rp + c^3 \lp a + a^6 + \dots \rp + c^4 \lp a^3 + a^8 + \dots \rp \bigg]. \nonumber \\
\end{eqnarray}
For general odd $n$ the generating function is then
\begin{equation}
F^{(\infty)}_S \lp a,c \rp = \left[ \sum_{m=0} ^{\infty} c^{nm} \right] \left[ \sum_{j=0}^{\frac{n-1}{2}} c^j \sum_{\ell = 0} ^{\infty} a^{n \ell + 2j} + c^{\frac{n+1}{2}} \sum_{j=0}^{\frac{n-3}{2}} c^j \sum_{\ell=0}^{\infty}a^{n \ell + 2j+1} \right].
\end{equation}
When we re-introduce the $B$ operators this becomes
\begin{equation}
\boxed{F^{(\infty)}_S \lp a,b,c \rp = \left[ \sum_{m=0} ^{\infty} c^{nm} \right] \left[ \sum_{j=0}^{\frac{n-1}{2}}  \sum_{\ell = 0} ^{\infty} \sum_{k=0}^{n \ell + 2j} a^{k}b^{n \ell + 2j - k}c^j + c^{\frac{n+1}{2}} \sum_{j=0}^{\frac{n-3}{2}}  \sum_{\ell=0}^{\infty} \sum_{k=0}^{n \ell + 2j + 1} a^{k} b^{n \ell + 2j+1-k}c^j \right]}.
\label{eq:fsc3znapp}
\end{equation}

If we make the replacements $a \rightarrow t$, $b \rightarrow t$, $c \rightarrow t$ in equation \eqref{eq:fsc3znapp} so that we only have a chemical potential for the $R$-charge then we get 
\begin{equation}
F^{(\infty)}_S \lp t \rp = \frac{-t^{2 n}-n t^{n+3}-t^{2 n+3}-2 t^{\frac{3 (n+1)}{2}}+2 t^{\frac{n+3}{2}}+n t^n+t^3+1}{\left(t^3-1\right)^2 \left(t^n-1\right)^2}.
\label{eq:fsc3zntapp}
\end{equation}
It can be shown that the result that can be obtained using the methods of \cite{Benvenuti:2006qr} for general odd $n$ matches equation \eqref{eq:fsc3zntapp}.

We can take the plethystic exponential to get the multi-trace operator generating function:
\begin{equation}
F^{(\infty)}_M(a,b,c) = \prod_{m=0}^{\infty} \left[ \prod_{j=0}^{\frac{n-1}{2}} \prod_{\ell=0}^{\infty} \prod_{k=0}^{n \ell + 2j} \frac{1}{1-c^{nm+j}a^k b^{n \ell + 2j - k}}\right]\left[ \prod_{j=0}^{\frac{n-3}{2}} \prod_{\ell=0}^{\infty} \prod_{k=0}^{n \ell + 2j + 1} \frac{1}{1-c^{nm+\frac{n+1}{2}+j}a^k b^{n \ell + 2j - k}} \right].
\end{equation}

Now, let's consider even $n$, again simplifying to the case with only $A$ and $C$ operators. For even $n$ we have to deal with the complication that not all operators with the same number of $C$'s will be equal, {\it e.g.,} $\Tr(C_{1,n-1} \dots C_{3,1}) \neq \Tr(C_{2,n}\dots C_{4,2})$. Operators with at least one $A$ in the trace will not have this feature. That is, all operators with $n_1$ $A$'s and $n_2$ $C$'s will be equal to one another when $n_1>0$. Our generating function is then
\begin{equation}
F^{(\infty)}_S(a,c) = \sum_{m=0}^{\infty} \lp c^{\frac{n}{2}}\rp ^m \sum_{\ell=0} ^{\infty} a^{n \ell} \sum_{j=0}^{\frac{n}{2}-1} c^j a^{2j} + \sum_{m=1}^{\infty} \lp c^{\frac{n}{2}} \rp ^m,
\end{equation}
which, when we re-introduce the $B$'s, becomes
\begin{equation}
\boxed{F^{(\infty)}_S (a,b,c) = \sum_{m=0} ^{\infty} c^{\frac{nm}{2}} \sum_{j=0}^{\frac{n}{2}-1} \sum_{\ell =0}^{\infty} \sum_{k=0}^{n \ell + 2j } a^{n \ell + 2j -k} b^k c^j + \sum_{m=1}^{\infty} c^{\frac{nm}{2}}}.
\label{eq:fsc3znevenapp}
\end{equation}
When we make the substitution $a \rightarrow t$, $b \rightarrow t$, $c \rightarrow t$ in equation \eqref{eq:fsc3znevenapp} we get
\begin{equation}
F^{(\infty)}_S(t) = \frac{\frac{n \left(1-\left(t^3\right)^{n/2}\right) t^n}{\left(1-t^3\right) \left(1-t^n\right)^2}+\frac{1-\left(t^3\right)^{n/2}}{\left(1-t^3\right) \left(1-t^n\right)}+\frac{2
   \left(-\frac{1}{2} n t^{3 n/2}+\left(\frac{n}{2}-1\right) t^{\frac{3 n}{2}+3}+t^3\right)}{\left(1-t^3\right)^2 \left(1-t^n\right)}}{1-t^{n/2}} + \frac{t^{n/2}}{1-t^{n/2}}.
\label{eq:fsc3zneventapp}
\end{equation}
It can be shown that the result that can be obtained using the methods of \cite{Benvenuti:2006qr} for general even $n$ is equal to the first term in equation \eqref{eq:fsc3zneventapp}. The second term accounts for the fact that there are two single-trace operators with $\frac{nm}{2}$ $C$'s for every $m \in \mathbb{Z}_+$.

If we take the plethystic exponential of this then we get the multi-trace operator generating function
\begin{equation}
F^{(\infty)}_M (a,b,c) = \left[\prod_{m=0} ^{\infty} \prod_{j=0}^{\frac{n}{2}-1} \prod_{\ell =0}^{\infty} \prod_{k=0}^{n \ell + 2j } \frac{1}{1-a^{n \ell + 2j -k} b^k c^{j+\frac{nm}{2}}}\right] \left[\prod_{m=1}^{\infty} \frac{1}{1-c^{\frac{nm}{2}}}\right].
\end{equation}

\bibliographystyle{JHEP}
\bibliography{bibliography}

\end{document}